\documentclass[twocolumn,times,tighten]{aastex631}
\usepackage[utf8]{inputenc}
\usepackage{amsmath,amstext,amssymb}
\usepackage{rotating}
\usepackage{booktabs}
\usepackage{numprint}
\usepackage{graphics}
\usepackage{graphicx, float}
\usepackage{epsfig}
\usepackage{wrapfig}
\usepackage{xcolor}
\usepackage[figure,figure*]{hypcap}
\usepackage{natbib}

\definecolor{Green}{rgb}{0,0.5,0}
\definecolor{Blue}{rgb}{0,0,1}
\definecolor{Red}{rgb}{1,0,0}

\newcommand{\kms}{km s$^{-1}$}

\newcommand{\msun}{M$_{\odot}$}

\newcommand{\HI}{\mbox{H\textsc{i}}}

\shorttitle{Infall Profiles for Supercluster-Scale Filaments}
\shortauthors{Odekon et al.}

\begin{document}

\title{Infall Profiles for Supercluster-Scale Filaments}

\author[0000-0003-0162-1012]{Mary Crone Odekon}
\affil{Department of Physics, Skidmore College, Saratoga Springs, NY 12866, USA; mcrone@skidmore.edu}
\author[0000-0002-5434-4904]{Michael G. Jones}
\affil{Steward Observatory, University of Arizona, 933 North Cherry Avenue, Rm. N204, Tucson, AZ 85721-0065, USA}
\author[0000-0001-8165-2323]{Lucas Graham}
\affil{Department of Physics, Washington University in St. Louis, St. Louis, MO 63105, USA; graham.l@wustl.edu}
\author[0000-0002-7727-1824]{Jessica Kelley-Derzon}
\affil{Department of Physics, University of Florida,Gainesville, FL, 32603, USA; jkelleyderzon@ufl.edu}
\author[0000-0001-9735-4232]{Evan Halstead} 
\affil{Department of Physics, Skidmore College, Saratoga Springs, NY 12866, USA; ehalstea@skidmore.edu}


\begin{abstract}
We present theoretical expectations for infall toward supercluster-scale cosmological filaments, motivated by the Arecibo Pisces-Perseus Supercluster Survey (APPSS) to map the velocity field around the Pisces-Perseus Supercluster (PPS) filament. We use a minimum spanning tree applied to dark matter halos the size of galaxy clusters to identify 236 large filaments within the Millennium simulation. Stacking the filaments along their principal axes, we determine a well-defined, sharp-peaked velocity profile function that can be expressed in terms of the maximum infall rate $V_{\rm max}$ and the distance $\rho_{\rm max}$ between the location of maximum infall and the principal axis of the filament. This simple, two-parameter functional form is surprisingly universal across a wide range of linear mass densities. $V_{\rm max}$ is positively correlated with the halo mass per length along the filament, and $\rho_{\rm max}$ is negatively correlated with the degree to which the halos are concentrated along the principal axis. We also assess an alternative, single parameter method using $V_{25}$, the infall rate at a distance of 25 Mpc from the axis of the filament. Filaments similar to the PPS have $V_{\rm max} = 612 \ \pm$ 116 km s$^{-1}$, $\rho_{\rm max} = 8.9 \pm  2.1$ Mpc, and $V_{25} =329 \ \pm$ 68 km s$^{-1}$. We create mock observations to model uncertainties associated with viewing angle, lack of three-dimensional velocity information, limited sample size, and distance uncertainties. Our results suggest that it would be especially useful to measure infall for a larger sample of filaments to test our predictions for the shape of the infall profile and the relationships among infall rates and filament properties. 

\end{abstract}


\section{Introduction}
\label{sec:intro}

As the initial density perturbations of the early universe grow, they form large-scale structures such as sheets, filaments and clusters. Observations of these various structures can provide tests of cosmological models over different spatial scales and in different dynamical environments.
For example, the observed infall of galaxies toward large-scale filaments can provide a test of cosmological models on scales intermediate between clusters of galaxies \citep[e.g.][]{sorce2018a} and larger bulk flows \citep[e.g.][]{howlett2022a}. The ongoing Arecibo Pisces-Perseus Supercluster Survey \citep[APPSS;][] {odonoghue2019a}, in particular, is designed to provide observational constraints on the infall rate to the main filament of the Pisces-Perseus Supercluster (PPS). Here we provide a theoretical comparison for observations of infall to the PPS and similar filaments.

The PPS was chosen as the target for APPSS in part because of its simple orientation geometry: it is oriented across the sky, approximately perpendicular to the line of sight. It is about 70 Mpc away, on the far side of the local void that separates the Virgo supercluster region from the PPS region \citep[see][]{haynes1986a}. 

The APPSS survey will determine galaxy distances using the Baryonic Tully-Fisher Relation \citep[BTFR;][]{mcgaugh2000a, papastergis2016a, lelli2016}, an observed correlation between galaxy rotation speed and baryonic mass that provides distance estimates to a precision of about 30\%.  By including all the baryonic mass in the galaxy -- gas mass as well as stellar mass -- the BTFR extends the traditional Tully-Fisher Relation \citep{tully1977a} down to low-mass galaxies, allowing us to adequately sample the volume surrounding the PPS. 

The APPSS survey is just one of a growing body of peculiar velocity catalogs used to test cosmological models. Especially notable are the recent Sloan Digital Sky Survey Peculiar Velocity Catalog \citep[34,059 objects;][]{howlett2022a}, the Cosmicflows-IV Tully-Fisher catalog \citep[9,800 objects;][]{kourkchi2020a}, and the 6-degree Field Galaxy Survey peculiar velocity sample \citep[8,800 objects;][]{springob2014a}. Large surveys like these, combined with smaller targeted surveys like APPSS that focus on specific regions, will provide data that can be compared with theoretical expectations using a variety of different techniques. For example, \citet{shaya2022a} use numerical reconstructions on objects from the Cosmicflows-3 catalog \citep{tully2016a} and groups from the Two Micron All Sky Survey \citep{tully2015a} to trace trajectories of galaxy systems in the local universe and constrain the value of the Hubble constant. 

An analytic treatment of the velocity field near filaments is more complicated than treatments of the velocity field on either smaller or larger scales. The regions around filaments are in an intermediate stage of evolution, neither dynamically relaxed like small, dense clusters nor in the linear regime. In addition, unlike infall toward clusters of galaxies, which can be approximated as spherical, the gravitational attraction of external sources along one direction does not cancel out for a cylindrical geometry. 
Therefore, rather than use an analytic approximation, we determine the expected infall using cosmological simulations. In particular, we use the Millennium simulation \citep{springel2005a} because it is large enough to provide a statistically useful sample of filaments. The Millennium volume is a periodic cube of length 500 $h^{-1}$ Mpc, or about 714 Mpc for h=0.7. As described below, we are able to obtain a sample of 236 large filaments in the Milliennium volume, including 40 with properties similar to the PPS. For comparison, the Illustris TNG300 \citep{pillepich2018a} simulation volume is a cube of length 300 Mpc, nearly 15 times smaller. 

The Millennium database also includes galaxy catalogs based on semianalytic models. To mimic the observational methods of APPSS, we trace infall using gas-rich galaxies from the catalog of \citet{lacey2016a}, which applies the GALFORM semianalytic model to the updated ``MR7" Millennium run.

Several previous studies have investigated the expected structure of velocity fields in the region of filaments. One particular approach is to measure the dispersion of galaxy redshifts within a filament, and relate this to the linear mass density of the filament \citep{eisenstein1997a, pereyra2020a}. This method has the significant advantage that it can be used for galaxies without redshift-independent distance measurements. For a filament aligned across the sky, like the PPS filament, it can be based entirely on the width of the filament in redshift space. This is somewhat analogous to determining the virial mass of a cluster of galaxies using the velocity dispersion.  However, it does not directly measure the mass in the same way as it would for a relaxed, virialized object. Within a filament, the velocity dispersion may be dominated by orbital motions of galaxies within embedded dense groups and clusters. In addition, the width of a filament in redshift space likely includes a large number of infalling galaxies as well as orbiting galaxies. Nonetheless, it is not surprising that the velocity dispersion of filaments may be {\it correlated} to their underlying mass density, because larger filaments are likely to have larger clusters and greater infall, and might therefore be used to estimate the underlying mass.  

Another reason to consider the velocity field of simulated filaments is its effect on galaxy evolution. For example, \citet{laigle2015a} and \citet{dubois2014a} find that the spins of halos embedded in filaments are aligned with the vorticity of the filament. Indeed, \citet{xia2021a} conclude that rotation is an important filament property, and \citet{wang2021a} find possible observational evidence for it.
Another example is presented by \citet{kraljic2019a}, who show that the properties of simulated galaxies at different redshifts support the picture of motion along filaments, away from saddle points between large nodes. 

Velocity fields can also be used as part of the criteria for defining filaments. While it is typical for filaments to be identified using the density field or some tracer thereof, simulations can take advantage of the full six-dimensional phase space information to define what is special about a filament.  \citet{buehlmann2019a}, for example, use the eigenvalues of the velocity dispersion tensor to identify filament-like environments where two dimensions are collapsed.   

In a study of filaments more closely related to the project we discuss in this paper, \citet{pereyra2020a} show that the average velocity field surrounding simulated filaments is perpendicular to the filament spine, while the motion within the filament (that is, close to the filament spine), is primarily along the filament. 

While these studies illustrate the importance of filament velocity fields in understanding cosmology and galaxy evolution, they are not designed to provide a precise and observationally practical comparison for projects like APPSS that measure the large-scale infall surrounding a large (supercluster-scale) filament. Our goal here is to provide expected infall rates as a function of observable properties of the filament, and to find an observationally feasible way to make these measurements.

In this context, we choose a methodology that is straightforward to apply to observations like those available for the PSS, and that focuses on measurable, large-scale features of the filament structure and surrounding velocity field. In particular, to compensate for observational uncertainties in distance, we average over many galaxies within a large region surrounding each filament. This means we are not sensitive to variations along the filaments, but only to the overall bulk flow toward or away from them. Our approach is to identify the primary, straight-line axis of each filament using principal component analysis, and then find the average motion toward or away from this axis for galaxies in a large volume surrounding it. This differs from studies that trace the curves of filaments on smaller scales and consider variations along them.

More generally, given the multi-scale nature of cosmological filaments, it is important to keep in mind that our study is specifically designed to address supercluster-scale filaments. The multi-scale nature of filaments can be seen in large-scale cosmological simulations \citep[e.g.,][]{kraljic2019a,galarraga-espinosa2020a} and observations \citep[e.g.,][]{carronduque2021a,pereyra2020a,tempel2016a}, and also in studies that focus on filaments that drain into particular clusters and superclusters, whether in simulations \citep[e.g.,][]{kuchner2020a,gouin2020a} or observations \citep[e.g.,][]{castignani2022a,einasto2020a,mahajan2018a}.

This paper is organized as follows. In Section 2 we describe our method for defining filaments. In Section 3 we present the results for infall profiles for the simulated filaments, including two different ways to parameterize infall, and show how the parameters relate to the observable structure of the filaments and the underlying dark matter distribution. In Section 4 we present results for mock observations, including measurement uncertainties, and in Section 5 we summarize our results and discuss the implications. 

Note that in Sections 2 and 3, we present distances and masses in terms of the dimensionless Hubble constant $h=H_o/100$ km s$^{-1}$ Mpc$^{-1}$, the same way distances and masses are presented in the Millennium simulation database.  For the mock observations in Section 4, which include $h$-dependent magnitude limits, we use $h=0.7$.
 
\section{Defining filaments}

Cosmological filaments are defined in a variety of ways, depending on the information available and the purpose of a particular study. If the information is from a cosmological simulation, for example, filament identification can be based on the dark matter distribution itself rather than the galaxy distribution. Methods for finding filaments are reviewed in \citet{libeskind2018a}, \citet{rost2020a}, and \citet{carronduque2021a}.

Our method for defining filaments is designed to be used in the context of galaxy surveys as well as simulations, and is similar to those of \citet{alpaslan2014a}, \citet{odekon2018a}, \citet{pereyra2020a}, \citet{bonnaire2020a}, and other graph-based methods \citep[see Section 2 of][]{libeskind2018a}.
We start with groups of galaxies with virial masses above $10^{14} \ h^{-1}$ \msun, roughly the mass range often referred to as galaxy clusters. We connect these large groups with a minimum spanning tree, and break the tree where connecting edges are greater than $17\ h^{-1}$ Mpc. The remaining structures that have at least six large groups are considered filaments. Breaking the tree at this particular edge length maximizes the number of filaments with at least six connected groups in the simulation.

Figure 1 illustrates this process applied to observations of the local universe, specifically the Two Micron All Sky Survey (2MASS) groups catalog of \citet{tully2015a}. This catalog has the advantage of covering nearly the entire sky, because it is based on infrared wavelengths that can penetrate through dust in the Milky Way down to low galactic latitudes. We apply the filament finder in three dimensions with each group placed at the distance cz/H$_o$. The left panel in Figure 1 shows all the groups above a virial mass of $10^{14}\  h^{-1}$ \msun \ out to cz = 10,000 km s$^{-1}$, within which the sample is approximately complete \citep[see Section 6 in][]{tully2015a}. The center panel in Figure 1 shows the resulting filaments with at least six connected groups, and with the PPS filament indicated by large blue dots. The right panel shows the positions on the sky of these large groups in the PPS filament in the context of the 2MASS galaxy distribution.

Using our filament identification process, the PPS filament is defined by eight large groups that have a combined virial mass of $2.9\times10^{15}\ h^{-1}$ \msun. If we define filament length as the straight-line distance between the two most distant groups, the PPS filament has a length of 38 $h ^{-1}$ Mpc. It is important to keep in mind, however, that different filament detection algorithms would result in different properties for the PPS and other filaments. Our approach is to use a method that works with existing observational data, and that succeeds in identifying the main body of the PPS filament.

\begin{figure*}
   \includegraphics*[width=\textwidth]{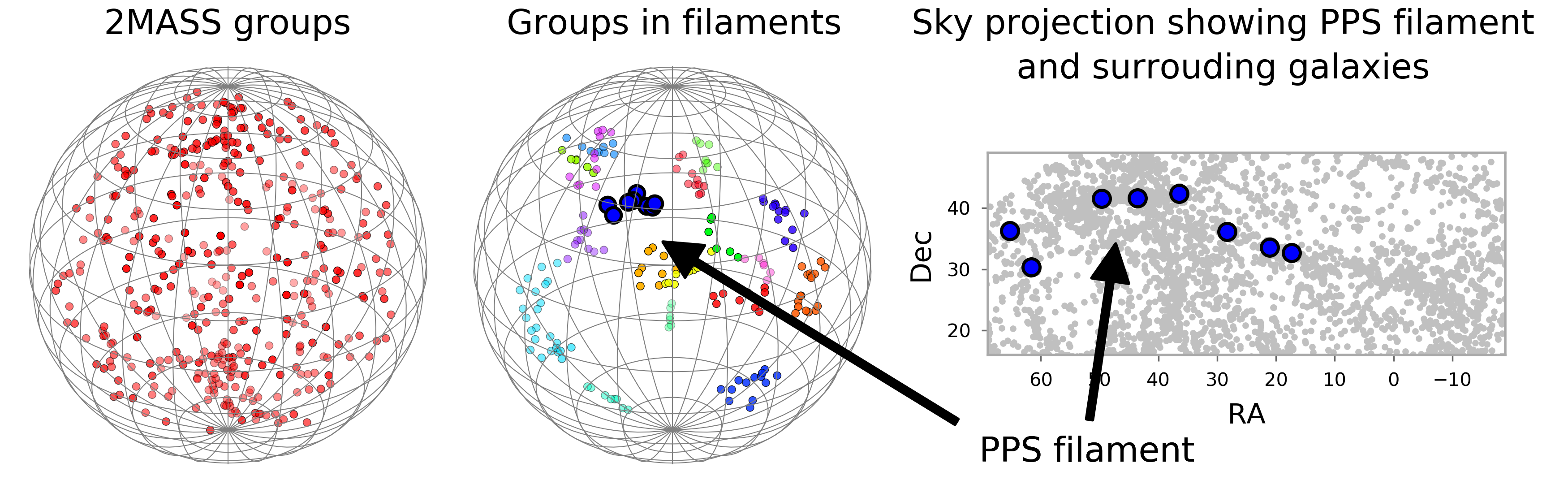}
    \caption{Results of our filament identification technique applied to the 2MASS groups catalog. The left panel shows all groups with $M_{vir}> 10^{14}h^{-1}$ \msun.  The center panel shows only groups identified to be in filaments, with different filaments indicated by color. The right panel is a sky view showing the groups in the Pisces-Perseus Supercluster filament (large blue dots) in the context of the 2MASS galaxy distribution (small gray dots).}
\end{figure*}

Figure 2 illustrates the filament identification process applied to the Millennium simulation. The left panel shows all halos with $M_{200} > 0.72\times 10^{14}h^{-1}$\msun, where $M_{200}$ is the mass within a region where the density is two hundred times the average density of the universe. This mass cutoff includes a correction factor of 0.72 to relate the $M_{200}$ halo masses in the Millennium database to the virial masses in the 2MASS group catalog \citep[see][]{duffy2008a, coe2010a, saez2016a}.
There are 5481 halos above this cutoff. The center panel in Figure 2 shows the resulting 236 filaments that have at least six halos, and the right panel shows one filament for which the elongated shape is clear in a y-z projection.

\begin{figure*}
   \includegraphics*[width=\textwidth]{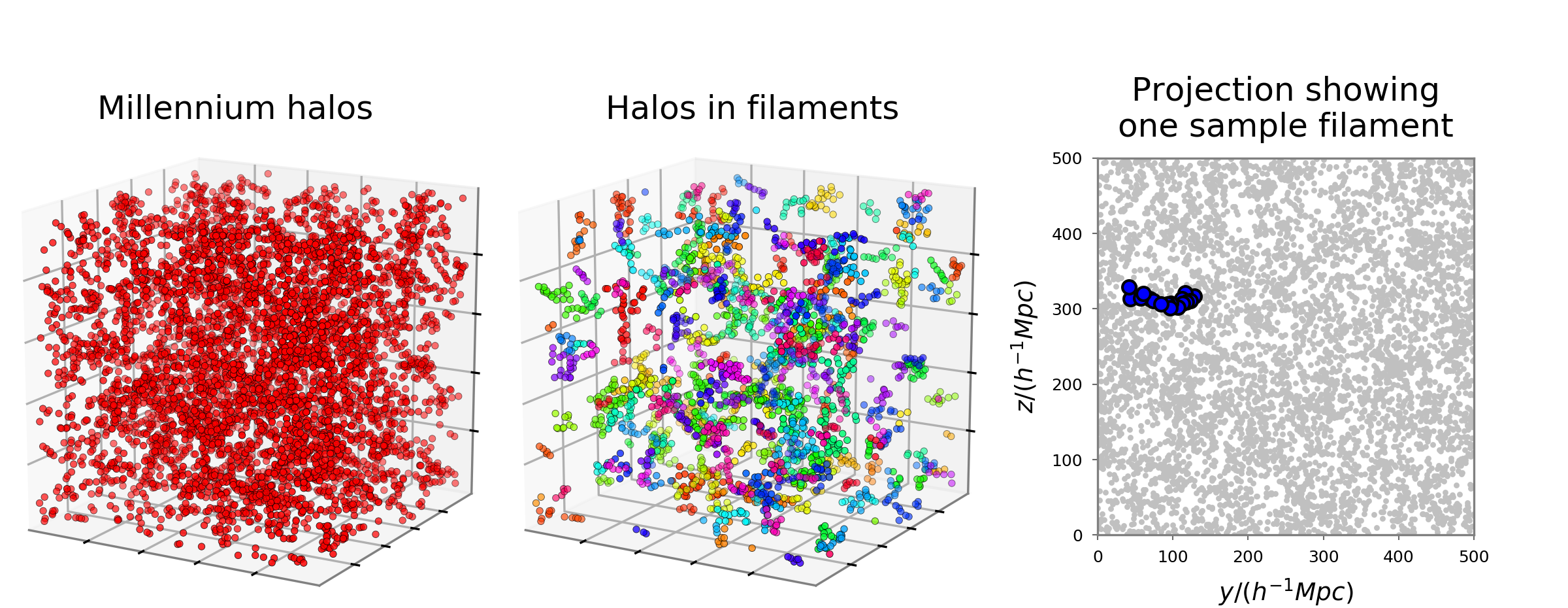}
    \caption{Results of our filament identification technique applied to the Millennium simulation. The left panel shows all halos with $M_{200}> 0.72 \times 10^{14}h^{-1}$ \msun, approximately equivalent to the cutoff of $M_{\rm vir}> 10^{14}h^{-1}$ \msun \ used in Figure 1 for the 2MASS groups. The center panel shows only halos identified to be in filaments, with different filaments indicated by color. The right panel is a rotated view showing the elongated geometry of a typical filament. Note that the Millennium simulation has periodic boundaries, so that some filaments are cut off on one side and continue on the opposite side.}
\end{figure*}

As illustrated in Figure 3, the lengths and combined group masses of the simulated filaments bracket those of the PPS filament, suggesting that these filaments are indeed an appropriate sample to compare with the PPS.

Figure 4 shows a histogram of the lengths of the simulated filaments, in a format similar to that used in several recent papers that describe filament catalogs. Note that different catalogs define filaments in different ways, resulting in different typical filament lengths. Our filaments tend to be larger, for example, than those in the three catalogs compared by \citet{rost2020a}, and in the lowest redshift block of \citet{carronduque2021a}. Our choice to define filaments using large groups is designed to detect supercluster-scale filaments like the PPS filament.

\begin{figure}
   \includegraphics[width=0.47\textwidth]{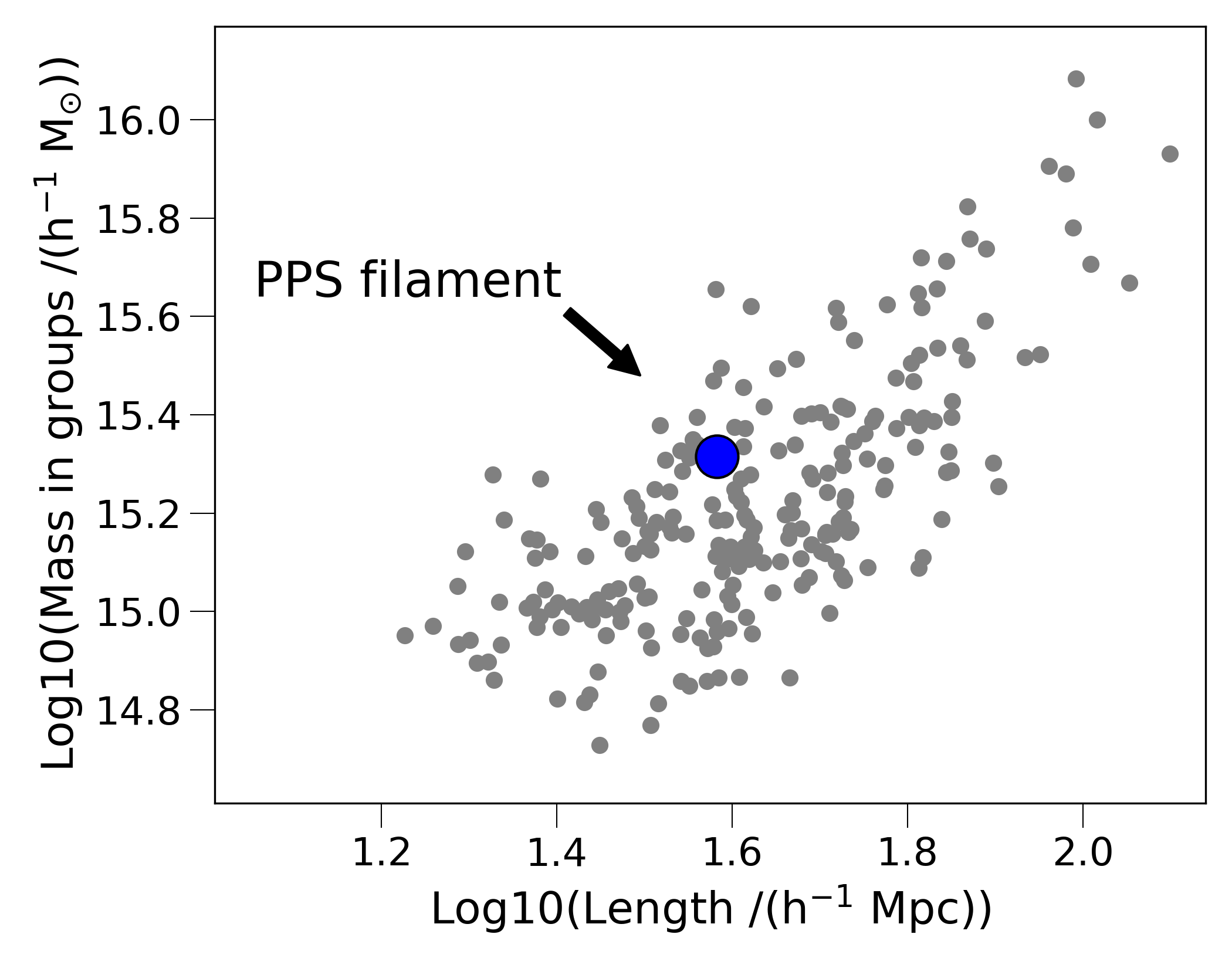}
    \caption{Comparison of the length and mass of the Pisces-Perseus Supercluster filament (large blue dot) with the simulated filaments (small gray dots). The horizontal axis shows the distance between the two most distant groups that define the filament. The vertical axis shows the sum of the virial masses of the groups or halos that define the filament. The range of values for the simulated filaments brackets the values for the PPS filament.}
\end{figure}

\begin{figure}
   \includegraphics[width=0.47\textwidth]{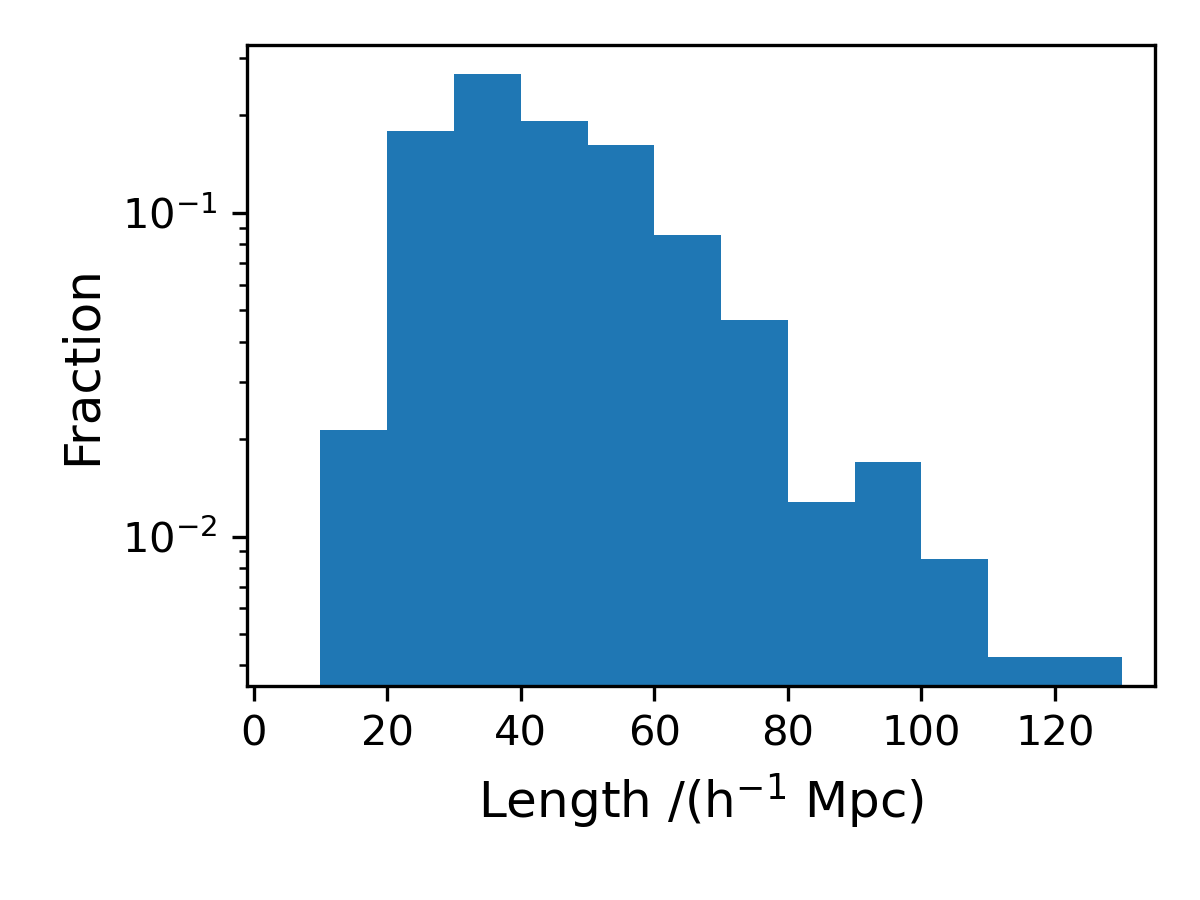}
    \caption{Histogram of filament lengths, in a format similar to Figure 19 of \citet{carronduque2021a}. Our filament detection algorithm is tuned to select larger filaments than some other filament catalogs \citep[e.g.][]{colberg2005a, rost2020a}.}
\end{figure}

\section{Determining Infall Profiles}

In this section we describe our method for selecting galaxies around the simulated filaments and measuring their infall rate as a function of distance from the filament axis.

The Millennium simulation \citep{springel2005a} uses a TreePM code to model gravity for collisionless matter, and then applies semi-analytical models to trace the formation and evolution of galaxies within merging dark matter halos. The Millennium database provides several simulated galaxy catalogs, some based on the original Millennium Run (MR), which used cosmological parameters consistent with 2dFGRS and first-year WMAP data, and others based on the updated MR7 \citep{guo2013a}, which used cosmological parameters consistent with the seven-year WMAP data. There are also two different semi-analytic codes used to create the Millennium galaxy catalogs. Some catalogs use the L-Galaxies code developed at the Max Planck Institute \citep{croton2006a,delucia2006a}, and some use the GALFORM code developed at the Durham University Institute for Computational Cosmology \citep{bower2006a}. In this paper, we focus on the \citet{lacey2016a} catalog, since it uses the updated MR7 cosmological parameters and the most recent semi-analytic models (which uses GALFORM). The APPSS survey is based on gas-rich late-type galaxies, so we make our sample selection based on cold gas mass, selecting galaxies with a minimum cold gas mass of $M_{min}=10^{8}h^{-1}$ \msun.  

For our purposes, galaxies serve mostly as tracers of the flow of matter toward filaments, so we do not expect our results to be very sensitive to the details of semi-analytic modeling. However, we expect the selection criteria for galaxies to make {\it some} difference, if only because early-type galaxies are more concentrated in groups and clusters. Therefore, their velocities will be more strongly affected by local orbits, potentially making it more difficult to measure the larger-scale flow toward filaments. 

We also make use of the galaxy catalog of \citet{lagos2012a}, which applies the GALFORM semi-analytic model to the earlier MR run. This is partly to check the dependence of our results on small changes to cosmological parameters. It also allows us to compare the galaxy velocity field to the large-scale underlying matter distribution, which is provided in the Millennium database for MR (but not MR7) in the form of a $256^3$ density grid. Note that the MR run is based on a different random realization of the initial conditions than the MR7, so it produces a totally different set of filaments. 

As described in the introduction, our goal is to determine an overall, large-scale infall profile for each filament. To do this, we select galaxies in a large cylindrical-shaped region centered on the main axis of each filament. We identify the axis by applying principal component analysis (PCA) to the large halos that were used to identify the filament, weighting the halos in proportion to their masses. The principal axis is defined as the straight line that minimizes the weighted distances of the halos from the axis. We then select the galaxies in a cylinder along the principal axis of each filament, with length 30 $h^{-1}$ Mpc along the principal axis of each filament, and radius that extends to 100 $h^{-1}$ Mpc. The length of 30 $h^{-1}$ Mpc is large enough to measure a representative infall profile, but small enough that it is dominated by the filament. The radius of 100 $h^{-1}$ Mpc extends far enough out to see the large-scale gravitational effect of the filament. 

We express infall in terms of cylindrical coordinates, with the axis of the coordinate system defined by the principal axis of each filament. We use the phrase ``infall profile" to mean V$_{infall}(\rho)$, where cylindrical coordinate $\rho$ is the perpendicular distance to the principal axis and the ``infall velocity" V$_{infall}= - \mathbf{v} \cdot \hat{\mathbf{\rho}}$ is the component of the velocity directed toward the principal axis (in the direction of negative $\rho$). 

Figure 5 illustrates the density and velocity structure for the cylinder-shaped region selected in this way surrounding a sample filament from the MR run. The top left panel shows the infall profile. Galaxies in the very center are primarily in virialized structures where orbital motions produce both positive and negative values for V$_\rho$. At about 10 $h^{-1}$ Mpc, nearly all galaxies are falling toward the filament, with a positive infall velocity. At large distances, galaxies are not on average affected by that particular filament, so the average infall speed approaches zero. Note that the values for velocity are in comoving coordinates, which do not include the expansion of the universe. 

The top right panel of Figure 5 uses the same color-coding based on infall velocity, but shows galaxy positions within the projected cylinder, with the principal axis directed perpendicular to the page. The dark blue region surrounding the center is the infall region.

The bottom row of Figure 5 is similar to the top row, but instead of showing infall it shows density. On the bottom left is the density for each grid point in the cylinder as a function of distance from the principal axis, and on the bottom right is an image of the projected cylinder, using the same color coding for density as in the bottom left. The region near the filament axis appears as a light yellow overdense region in the center of the cylinder surrounded by a complicated web of smaller filamentary structures.  

\begin{figure*}
   \includegraphics*[width=1.00\textwidth]{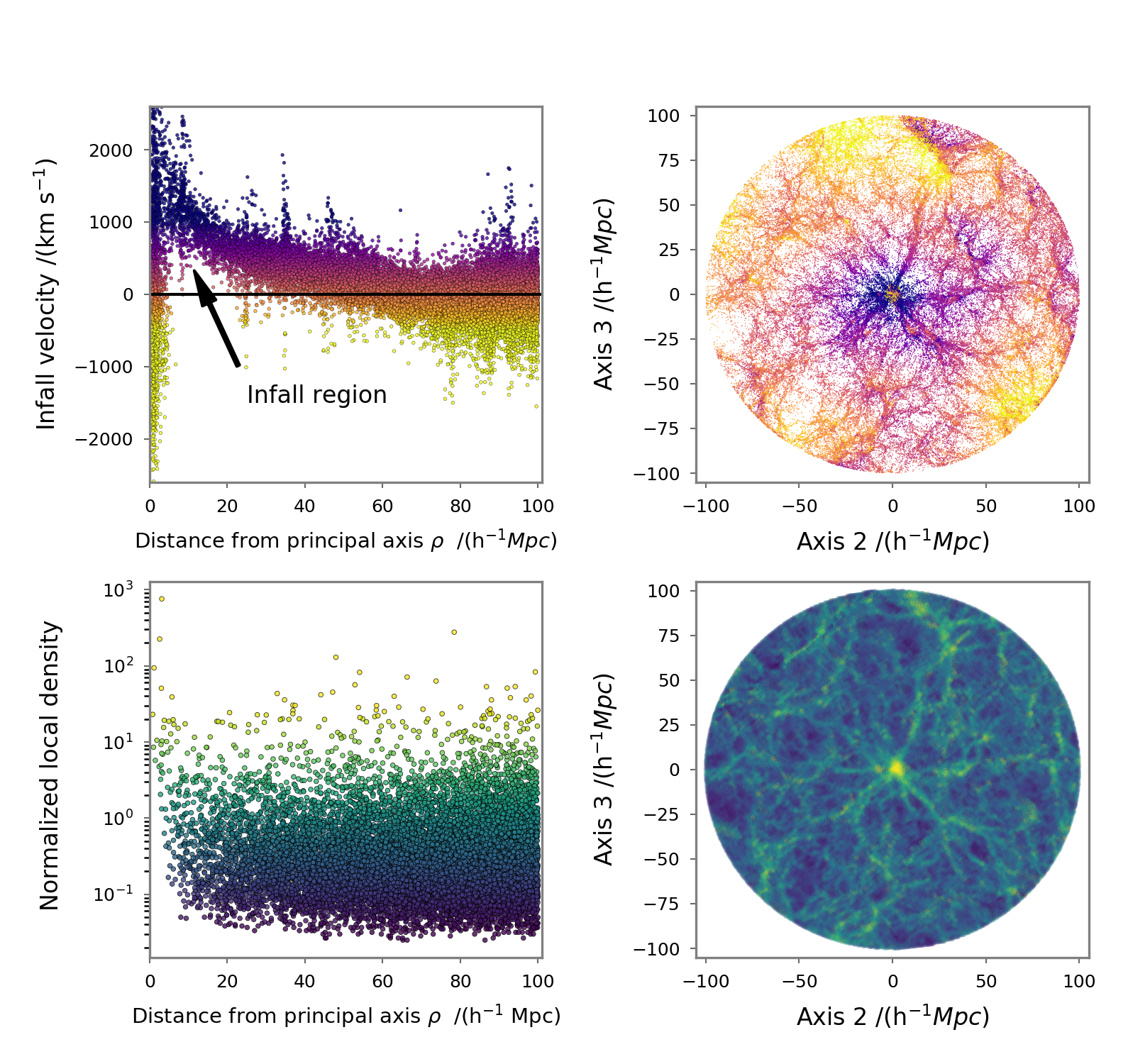}
    \caption{The velocity and density structure in a cylinder-shaped region of thickness 30 h$^{-1}$ Mpc surrounding one sample simulated filament. The top left panel shows the component of velocity toward the principal axis of the filament as a function of distance $\rho$ from the principal axis. The top right panel shows the cylinder-shaped region projected along the principal axis, with the same color coding for infall velocity as on the left. Galaxies in the very center are primarily in virialized structures where orbital motions produce both positive and negative infall velocities, while at about 10 $h^{-1}$  Mpc, nearly all galaxies are falling toward the filament (positive infall velocity). At high distances from the filament, the infall speed approaches zero. (Note that the values for velocity are in comoving coordinates which do not include the expansion of the universe.) The bottom left panel shows density as a function of distance from the principal axis, and the bottom right panel shows the projected cylinder with the same color coding for density as on the left. The filament appears as a light yellow dense region in the center surrounded by smaller filamentary structures.}
\end{figure*}

The cylinders we examine around each filament are quite large, so we can examine the shape of the infall profile out to large distances. To illustrate the scale, as well as the geometry of the cylinder and principle axis, Figure 6 shows the cylinders surrounding three of the filaments in the context of the entire Millennium volume. The groups used to define the filaments are indicated with large blue dots, and the galaxies are indicated with small points that are color-coded for infall velocity as in Figure 5. The orientation of the cylinders is defined relative to the principle axis of each filament. 

\begin{figure}
   \includegraphics[width=0.49\textwidth]{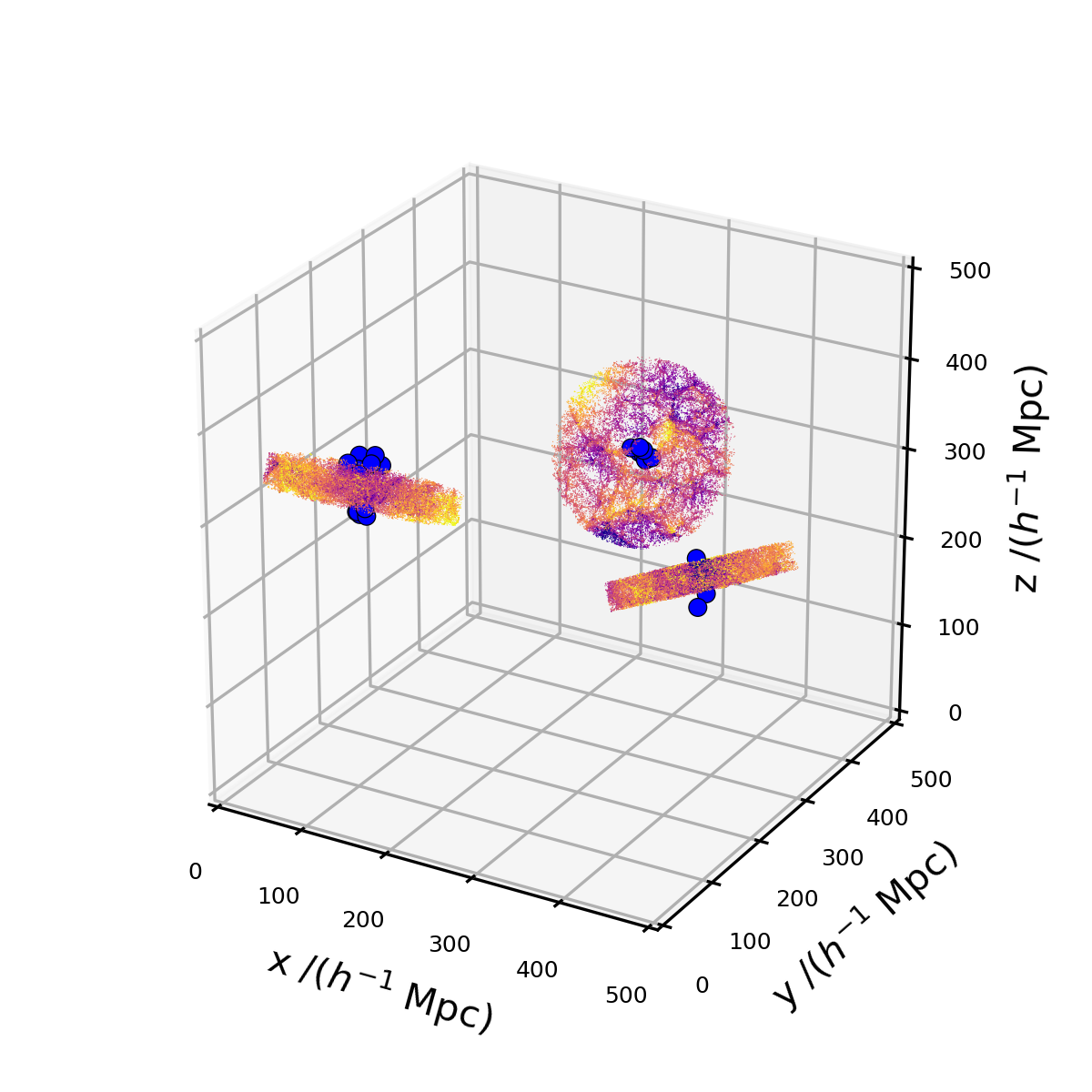}
    \caption{Cylinders perpendicular to the principal axis for three sample filaments,showing the scale of the regions from which we draw galaxies to measure the infall profile. Large blue dots are the groups that define each filament. Galaxies in the cylinders are color-coded for infall velocity in the same way as the projected cylinders in the upper right of Figure 5.}
\end{figure}

To find an average infall profile function, we average together the 236 simulated filaments in the MR7 run by stacking them along their principal axes. The resulting infall profile, if we average the galaxy infall velocities in bins of distance from the principal axis, is shown in the left panel of Figure 7. Each filament is equally weighted in each distance bin; the average profile exhibits considerably more scatter if we instead weight each {\it galaxy} the same, so that the profile is more heavily weighted by whichever filaments have the most galaxies at that particular distance from the principal axis. Note that the infall profiles are calculated as an average in shells (binning by the distance $\rho$), not as an integrated average of all galaxies within $\rho$.

\begin{figure}
   \includegraphics*[width=0.49\textwidth]{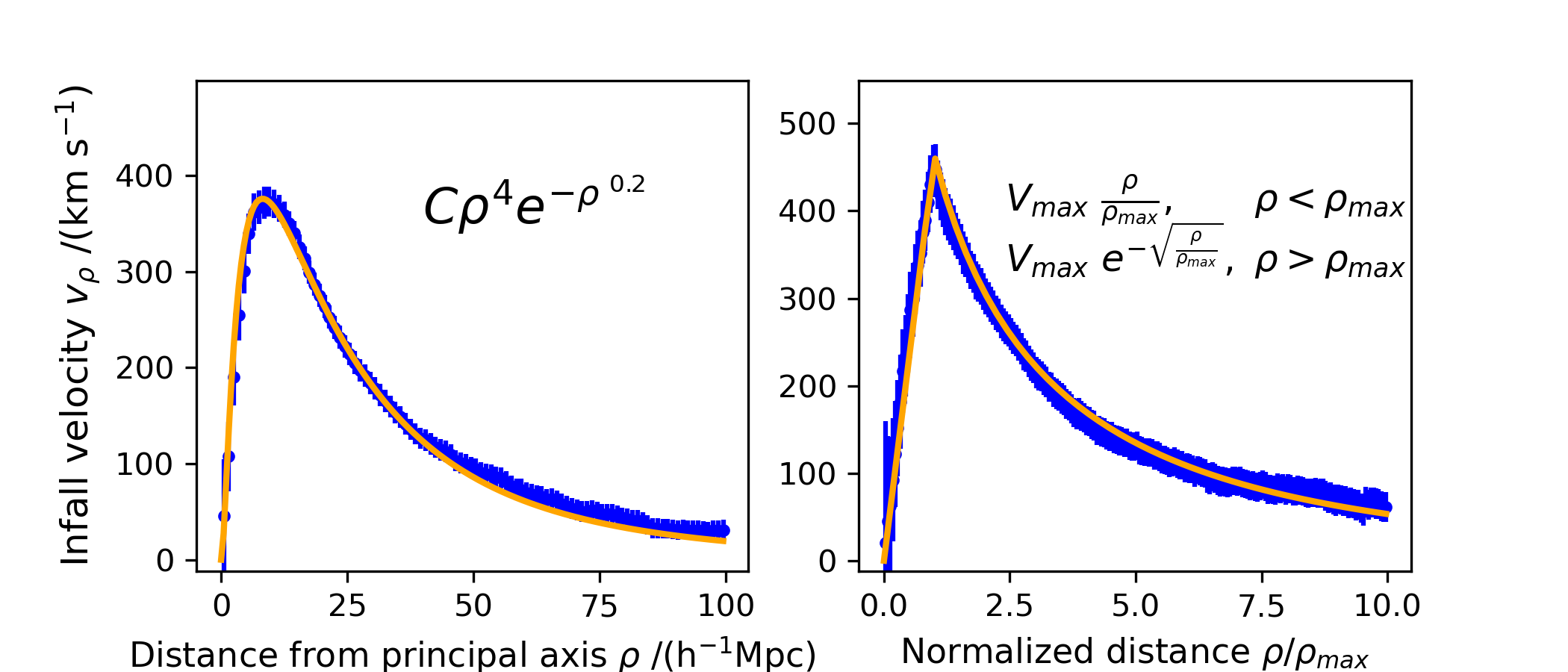}
    \caption{Average velocity profiles for all 236 simulated filaments from the MR7 run. On the left, infall velocities are stacked by distance from the principal axis. On the right, the distance for each filament is first normalized by its peak infall distance before stacking. Blue dots and error bars show binned means and the standard error in the mean. Orange lines show the functions given in the upper right of each panel. The peak in the left panel is rounded because different filaments peak at different distances. For a single filament, it is more appropriate to use the sharp-peaked function in the right panel.}
\end{figure}

A simple function that can be used to describe the shape in the left panel of Figure 7 is
\begin{equation}
v = C\rho^4 e^{{-\rho}^{ \ 0.2}}
\end{equation}
where $\rho$ is the distance from the principal axis. The rounded peak is the result of averaging together filaments with velocity profiles that peak at different distances. To find a function that represents the typical infall profile for a {\it single} filament, it is more appropriate to normalize the distances by the peak infall distance before averaging. To do this, we first average the infall velocities of the galaxies for each filament in distance bins of width 1 Mpc $h^{-1}$ and find the distance bin with the highest average infall. The distances from the principal axis are then divided by this distance before stacking together with the other filaments. The right panel of Figure 7 shows the resulting average profile, which we fit to the function 

\begin{equation}
\label{eqn:2paraminfall}
\begin{split}
v = V_{\rm max}\  \frac{\rho}{\rho_{\rm max}}, \ \ \ \ \ \ \rho \leq \rho_{\rm max} \\
v = V_{\rm max}\ e^{-\sqrt{\frac{\rho}{\rho_{\rm max}}}},  \ \rho \geq \rho_{\rm max}.   
\end{split}
\end{equation}

For the averaged profile in the right panel of Figure 7, the peak occurs by construction at $\rho/\rho_{max} = 1$.  But for an individual filament, Equation \ref{eqn:2paraminfall} can be fit to the infall profile in terms of $\rho$ to determine the two parameters $V_{\rm max}$ and $\rho_{\rm max}$. 

Figure 8 shows this function fit to fifteen of the 236 MR7 filaments. The infall profiles show a wide variety of shapes due to the range of internal structure and sourrounding environment for each filament. Several filaments are in the vicinity of large structures that create clear gravitational signatures in the outer parts of these profiles. Filaments 3 and 6, for example, have structures at a distance of about 100 Mpc that cause the profile to be lower on the close side (smaller distance) and higher on the far side. Filament 14 has a large structure more than 100 Mpc away that pulls the entire outer part of the profile to lower infall velocities. Nonetheless, the fitting function reliably finds the peak in the infall profile. 

\begin{figure*}
   \includegraphics*[width=1.05\textwidth]{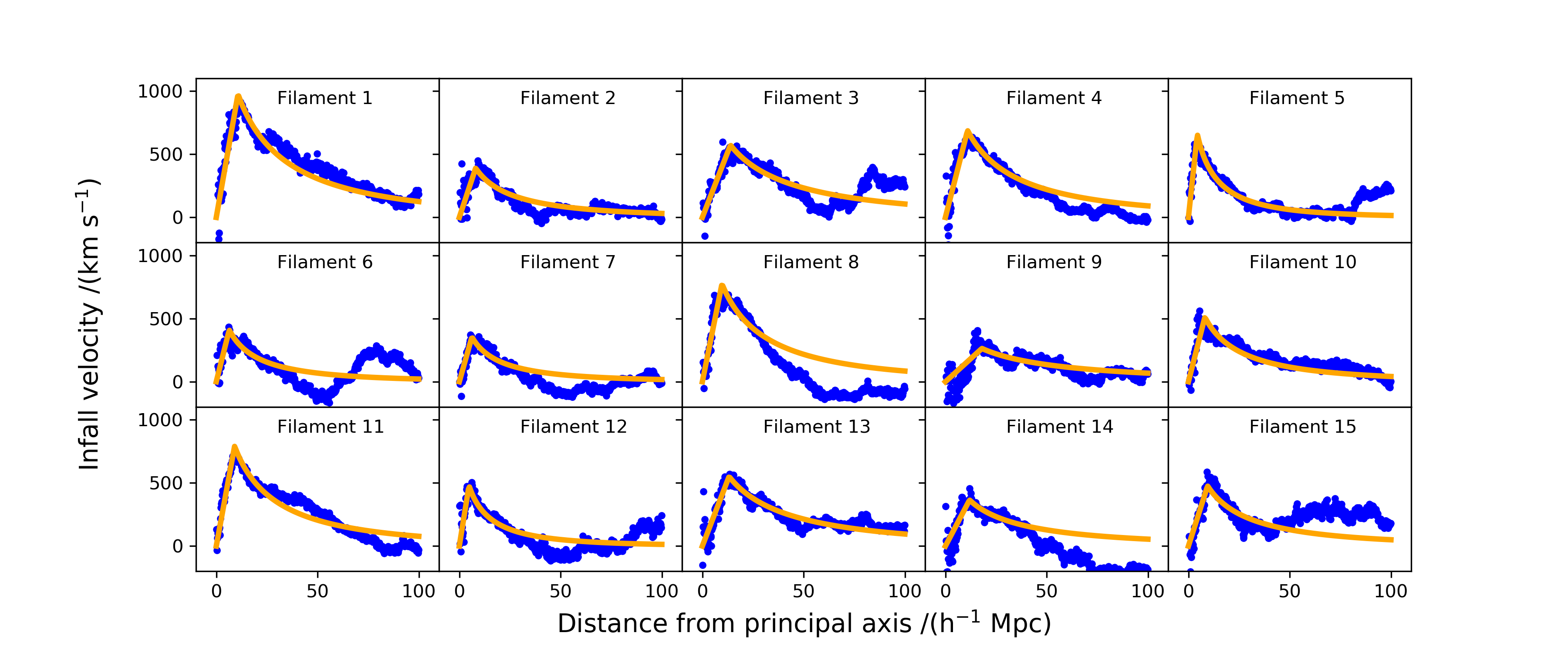}
    \caption{Velocity profiles as a function of distance from the principal axis for fifteen of the 236 simulated filaments. Blue dots show averages in bins of 1 $h^{-1}$ Mpc. Orange lines show the fitting function in Equation 2 applied to each filament. Filaments vary based on internal structure and surrounding environment, but the fitting function typically succeeds in identifying the maximum infall velocity and its distance from the principal axis.}
\end{figure*}

\subsection{A Two-Parameter Description of Infall: $V_{\rm max}$ and $\rho_{\rm max}$}

In this subsection we show that the parameters $V_{\rm max}$ and $\rho_{\rm max}$, as determined using Equation \ref{eqn:2paraminfall}, are related to two different aspects of the internal structure of the filaments: the overall mass of the filament, and the degree to which the mass is concentrated close to the principal axis. These two aspects of the structure can be seen in either the positions and masses of the groups used to identify the filaments, or in the full underlying mass distribution, including dark matter. 

First, we note that $V_{\rm max}$ and $\rho_{\rm max}$ are independent parameters, both required to specify the infall profiles for a particular filament (Figure 9). If the situation were simpler, and infall profiles for different filaments could be differentiated by a single parameter -- for example, a parameter related to the overall size of the filament -- we would expect to see a correlation between $V_{\rm max}$ and $\rho_{\rm max}$, indicating that they could be reduced to one parameter. 

The top row of Figure 10 shows how $V_{\rm max}$ and $\rho_{\rm max}$ are related to two different properties of the filament: the mass per length of large groups along the filament, and the degree to which the groups are concentrated toward the principal axis. The groups used to define these properties are the same ones used to define the filaments -- those with masses greater than $M_{200}> 0.72 \times 10^{14}h^{-1}$ \msun \ and not separated by more than the cutoff linking length of  $17\ h^{-1}$ Mpc.  The mass per length is defined as the sum of all the group masses in the filament divided by the straight-line distance between the two most distant groups. These are the same definitions for mass and length used in Figure 3. The group concentration is defined as the sum of the masses of groups within 5 h$^{-1}$ Mpc of the principal axis divided by the sum of the masses in groups within 15 h$^{-1}$ Mpc. {These groups are the same ones used to define the filaments and to calculate mass and length. This specific definition of groups concentration is somewhat arbitrary, but is useful because it produces a wide range of values when applied to  the simulated filaments.

Positions and masses of groups are observable properties that can provide predictions for the expected values of $V_{\rm max}$ and $\rho_{\rm max}$ for a specific filament like the PPS. Figure 10 shows the values for the PPS as vertical gray lines. If we estimate the uncertainty in mass per length for the PPS to be 30\% (e.g. by adding in quadrature $\sim 20\%$ uncertainty in mass to light ratio and $\sim 20\%$ in the amount of light \citep{tully2015a}), then the PPS has group mass per length in the range 5625 $\pm$ 1591 M$_\odot$ Mpc$^{-1}$. If we estimate the group concentration to be anywhere in the range 0.6-1.0, then forty of the simulated filaments are ``like the PPS" in terms of their group mass per length and group concentration. These forty filaments have an average $V_{\rm max}=612$ km s$^{-1}$ with standard deviation 116 km s$^{-1}$, and an average $\rho_{\rm max} =6.2$ h$^{-1}$ Mpc with standard deviation 1.5 h$^{-1}$ Mpc.

The bottom row of Figure 10 shows how $V_{\rm max}$ and $\rho_{\rm max}$ are related to the underlying mass distribution, including dark matter. Unlike the group structure, the underlying mass distribution is not directly observable. Instead, it can be inferred from measurements of $V_{\rm max}$ and $\rho_{\rm max}$ assuming the cosmological models are correct. In order to compare infall with the matter distribution, we use filaments from the MR Millennium run. (As described above, the MR has a density grid available in the Millennium database.)  The overdensity in the bottom left panel is defined as the density within a cylinder of radius 5 h$^{-1}$ Mpc and length 30 h$^{-1}$ Mpc oriented along the principle minus the average density of the universe, normalized by the average density of the universe. The mass concentration in the bottom right panel is defined as the mass within a cylinder of radius 5 h$^{-1}$ Mpc and length 30 h$^{-1}$ Mpc oriented along the principle axis, divided by the mass within a cylinder of radius 15 h$^{-1}$ Mpc and length 30 h$^{-1}$ Mpc. As was the case for the specific definition of group concentration, these definitions are  somewhat arbitrary, but are useful in that they produce a fairly wide range of values for the simulated filaments.

\begin{figure}
   \includegraphics[width=0.43\textwidth]{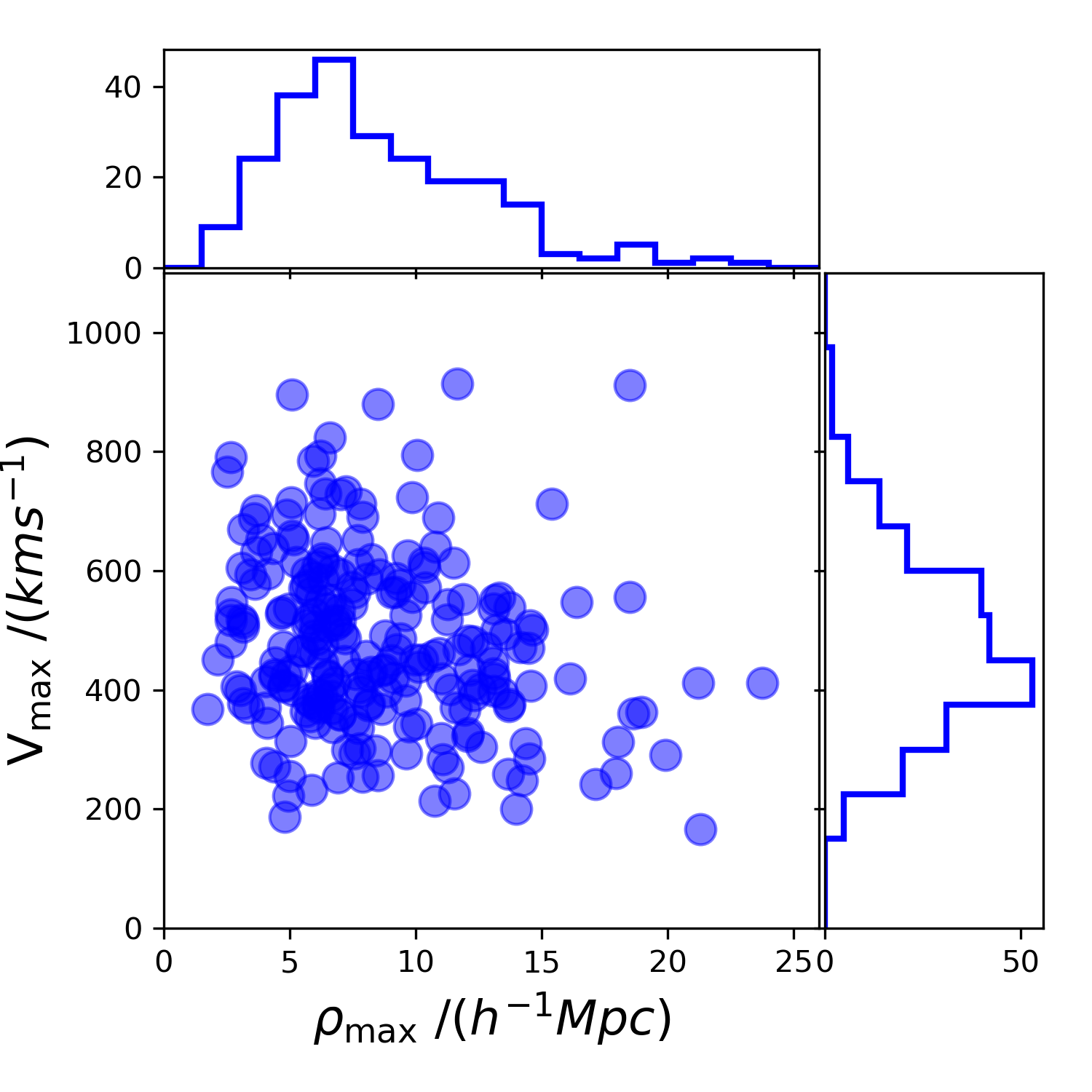}
    \caption{Distribution of values for $V_{\rm max}$ and $\rho_{\rm max}$ for the simulated filaments. These two parameters are independent and reflect different aspects of filament structure; they are not correlated as one would expect if the infall profile depended on a single underlying parameter.}
\end{figure}

\begin{figure*}
   \includegraphics*[width=1.0\textwidth]{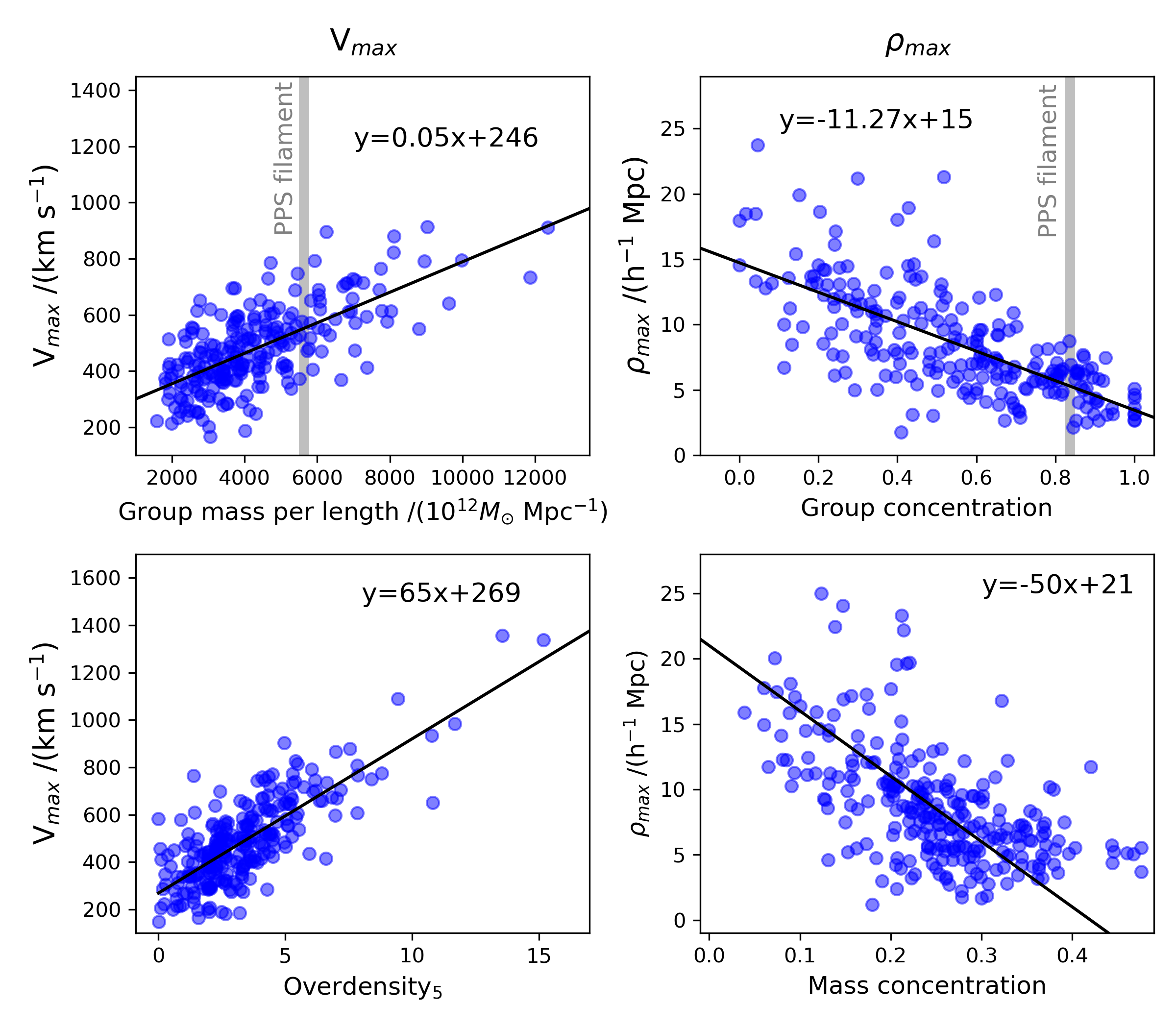}
    \caption{Relationships between $V_{\rm max}$ (left) and $\rho_{\rm max}$ (right) and the underlying mass distribution in of each filament. The upper panels show relationships with the mass distribution of the groups used to define the filaments:  $V_{\rm max}$ is positively correlated with the mass of the groups in the filament divided by the length of the filament, and  $\rho_{\rm max}$ is negatively correlated with degree to which groups are concentrated toward the principal axis of the filament. Vertical gray lines show the observed values of group mass per length and group concentration for the PPS filament, from which we can predict the expected values of $V_{\rm max}$ and $\rho_{\rm max}$.  The lower panels show relationships with the full mass distribution, including dark matter. The dark matter distribution is not directly observable, but could instead be inferred from measured values of $V_{\rm max}$ and $\rho_{\rm max}$. See the text for the specific definitions of group concentration, mass concentration, and overdensity.}
\end{figure*}

\subsection{A One-Parameter Description of Infall: $V_{25}$}

In this subsection we provide an alternative, one-parameter, method of using Equation 2 to predict the infall rate to filaments, and to use the observed infall rate to infer the underlying dark matter distribution. This method provides less information about the filament than the two-parameter description in terms of $V_{\rm max}$ and $\rho_{\rm max}$. However, it is more robust to observational limitations. 

As we describe in section 4 below, when we take into account observational limitations it becomes more difficult to reliably determine $V_{\rm max}$ and $\rho_{\rm max}$ separately. However, the value of the infall velocity at a fixed distance of 25 Mpc can still be recovered, and this value is also related to the structure of the filament. (Note that the value of $V_{\rm 25}$ is based on a distance of 25 Mpc assuming h=0.7, not a distance of 25 h$^{-1}$ Mpc.)

As shown in Figure 11, $V_{\rm 25}$ is related to both the group mass and to the overdensity of the filament. Here we define group mass as the mass of the groups used to define the filament, when limited to the central 30 h$^{-1}$ Mpc along the principle axis. In the lower panel, we define overdensity in terms of the mass within a cylinder of radius 15 h$^{-1}$ Mpc and length 30 h$^{-1}$ Mpc. The value of group mass, when defined in this way, is 2.0 h$^{-1} 10^{15} M_\odot$, or in the range 1.4-2.6 h$^{-1} 10^{15}$ for a mass uncertainty of 30\%. The subset of the 40 filaments that are ``like the PPS" as defined above, and that are in this range have an average $V_{\rm 25}=329$ km s$^{-1}$ with standard deviation 68 km s$^{-1}$.

The relationships for $V_{\rm 25}$ in Figure 11 are generally more tightly correlated than those for the more complicated two-parameter description in Figure 10. The relationship in the upper panel of Figure 11 has an adjusted coefficient of determination $R^2_{adj}= 0.638$, and the lower panel has $R^2_{adj}= 0.764$. For comparison, the relationships for $V_{\rm max}$ in Figure 10 have $R^2_{adj}= 0.456$ (upper panel) and $R^2_{adj}= 0.604$ (lower panel), and the relationships for $\rho_{\rm max}$ have $R^2_{adj}= 0.488$ (upper panel) and $R^2_{adj}= 0.323$ (lower panel).

\begin{figure}
   \includegraphics*[width=0.47\textwidth]{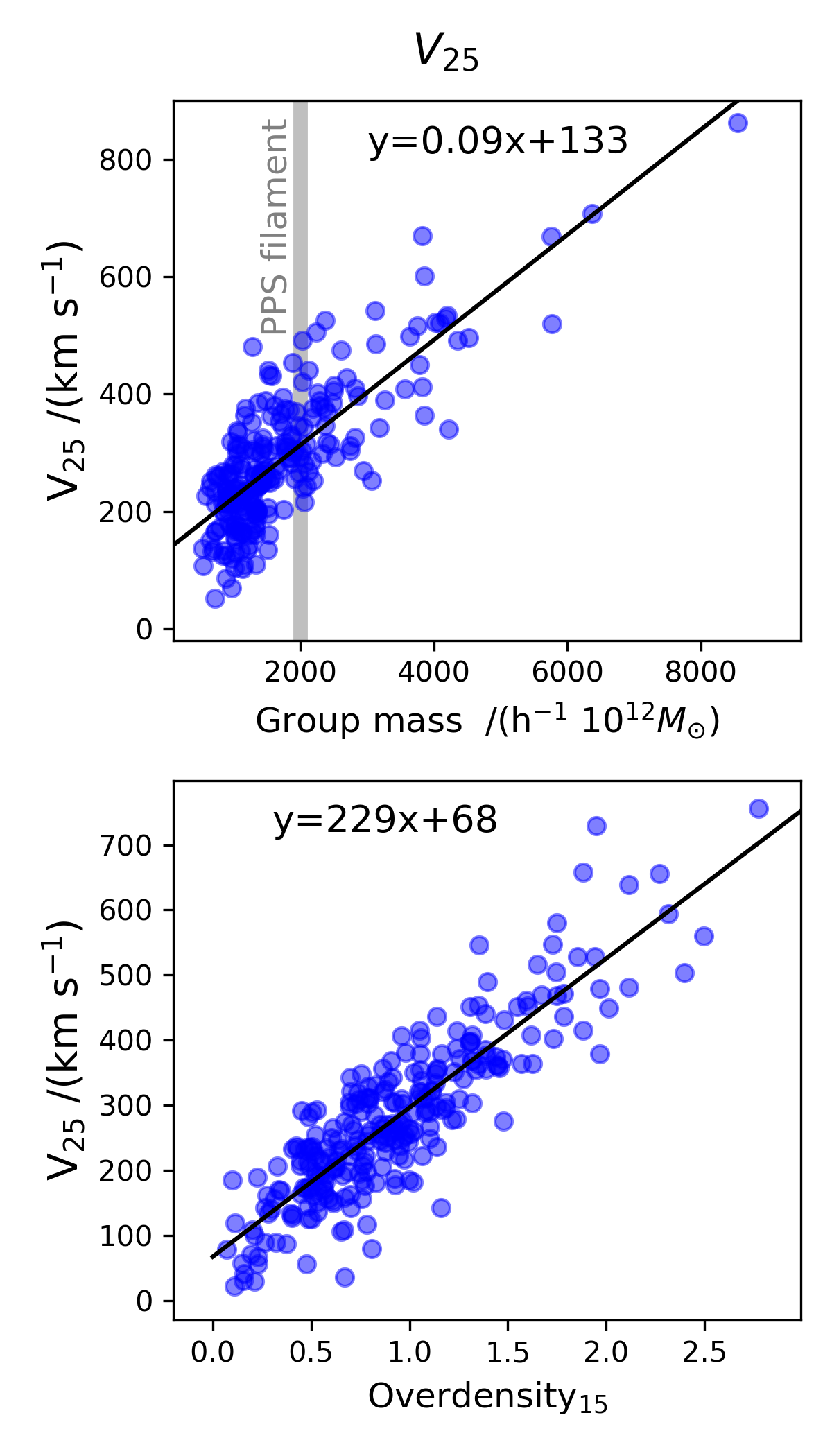}
    \caption{Similar to Figure 10, but for the description of infall in terms of the single parameter $V_{25}$, the value of the infall velocity at a fixed distance of 25 Mpc. $V_{25}$ is correlated with the total mass of the groups used to define the filament, and with the underlying overdensity of the filament.}
\end{figure}

When relating $V_{\rm max}$, $\rho_{\rm max}$, and $V_{\rm 25}$ to the mass distribution of the filament, we fit to a straight line $y(x) = mx+b$ rather than assuming a different functional form. We investigated fitting instead to a power law $y(x) = ax^m$, but found that the straight line was generally a better fit in the sense of having a higher$R^2_{adj}$. There are physical reasons one might expect a different functional form (for example, both $V_{\rm max}$ and $V_{\rm 25}$ going to zero as the overdensity goes to zero), but the scatter is such that we cannot distinguish among more complicated models.

Before discussing mock observations, we pause to comment on the dependence of our results on cosmological parameters and galaxy selection criteria. We can compare the results in the upper rows of Figures 10 and 11, which are based on the MR7 run and for which galaxies were selected by cold gas mass, with the same analysis performed on the MR run, or with galaxies selected by stellar mass. In all cases, the best-fit lines are consistent with those shown in Figures 10 and 11, so our results are robust to these differences.  For example, a least-squares fit to the relationship between $V_{\rm max}$ and group mass per length (upper left of Figure 10) yields a slope of $0.054\pm0.004$ and intercept of $246\pm18$.  If we instead select the galaxies by stellar mass, we find a slope of $0.053\pm0.004$ and intercept of $226\pm18$, and if we instead use the values from the MR run (which uses a different random realization of the initial conditions, as well as slightly different cosmological parameters), we find a slope of $0.045\pm0.004$ and intercept of $249\pm17$.  Similarly, the distribution of values of $V_{\rm max}$, $\rho_{\rm max}$, and $V_{25}$ for the MR7, cold-gas selected sample are not statistically distinguishable from the distributions of the same quantities obtained from the MR run or the stellar-mass selected galaxies.  For example, the p-value for the two-sided Anderson-Darling test is greater than 0.10 for each of these distributions as compared with the MR7, cold-gas selected sample.

\section{Mock Observations}

The previous sections provide a description of infall toward filaments based on the full information available in simulations. In this section, we discuss the feasibility of measuring infall to filaments given observational limitations similar to those of the APPSS survey, in particular:
\begin{itemize}
\item{\bf Viewing angle.} Even if we assume the filament is oriented across the sky, it will not look the same from all viewing directions unless the filament and the region surrounding it are axially symmetric. Here, we quantify the variation caused by lack of axial symmetry, when the infall profile is determined from galaxies limited to a region similar to that of APPSS.

\item {\bf Lack of three-dimensional velocity information.} For the simulations, we have three-dimensional information for both position and velocity.  For the observations, we have three position coordinates (Right Ascension, Declination, and distance) and one coordinate that provides information about velocity, the line of sight redshift. The redshift information, combined with distance (and the Hubble constant), gives peculiar motion along the line of sight. The observational coordinates can be converted into distance from the filament $\rho$ and motion toward the filament $v_\rho$, but converting line of sight velocity to motion toward the filament requires assumptions about the three-dimensional direction of motion.

\item {\bf Limited sample size.} An observational sample will be limited to the galaxies that reside within the survey volume, and the realities of a flux-limited observational sample will mean that fewer low-mass objects are included at greater distances. 

\item {\bf Measurement uncertainties.} While the measurement uncertainties for redshift and sky position are small, the uncertainty for distance may be as high as $\sim$30\%. Furthermore, with BTFR distance estimates the distance uncertainties are typically log-normal in shape, with long tails extending to higher distances.

\end{itemize}

\begin{figure*}
   \includegraphics*[width=1.0\textwidth]{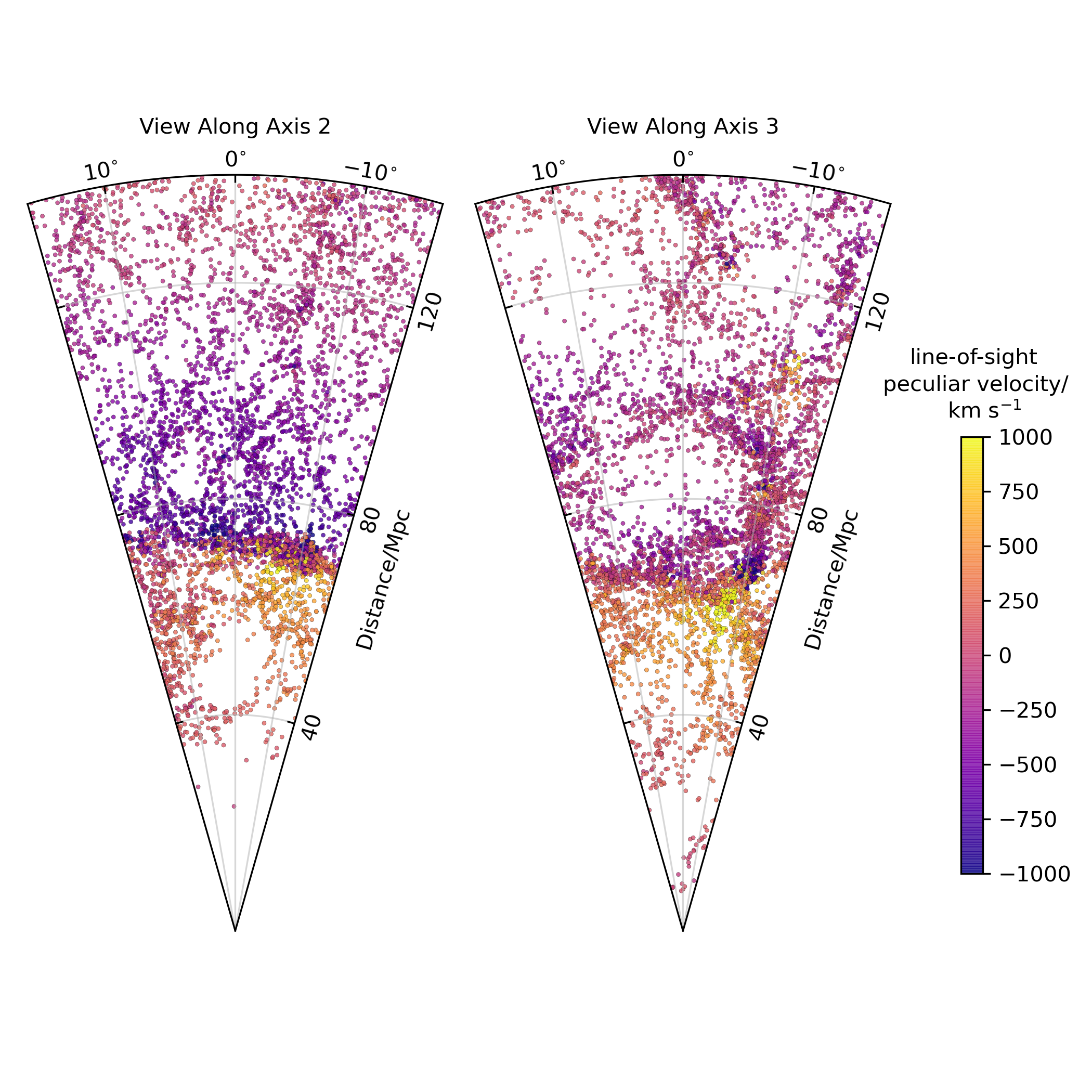}
   \vspace{-0.9in}
    \caption{Mock observational wedges for a single filament as seen from two different directions. In both cases, the principal axis of the filament is oriented across the field of view at a distance of 70 Mpc. For the wedge on the left, the observer's line of sight is along axis 2, and on the right the view is rotated by 90 degrees, along axis 3. Each galaxy is color-coded by the component of the peculiar velocity that is along the line of sight. Because the filaments and their surroundings are not axisymmetric, infall looks different from the two different perspectives.}
\end{figure*}

\begin{figure*}
   \includegraphics*[width=1.0\textwidth]{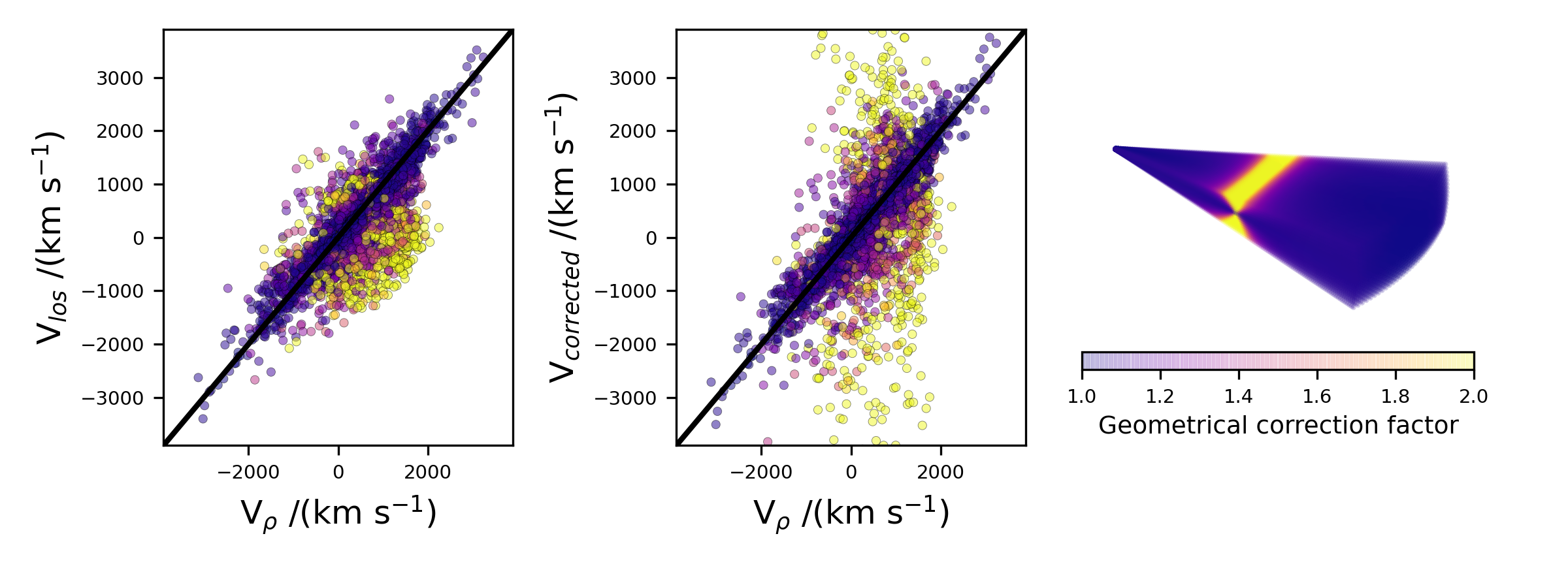}
    \caption{The impact of limited velocity information in the mock observations. The left panel shows the relationship between the measurable line-of-sight velocity component $V_{los}$ and the infall velocity $V_{\rho}$ for galaxies in an observational wedge around a single simulated filament. The central panel is similar, but with $V_{los}$ modified by the geometrical correction factor that would convert it to $V_{\rho}$ if the velocity is in the direction of the filament. The right panel shows the geometry of the observational wedge. All three panels are color-coded by the geometrical correction factor. For galaxies in most parts of the observational wedge (blue), $V_{los}$ and $V_{corrected}$ are both similar to $V_{\rho}$. However, for galaxies where the line of sight is nearly perpendicular to the direction of infall (yellow), it is difficult to recover the infall velocity, and applying the correction factor introduces additional scatter.}
\end{figure*}

\begin{figure}
    \centering
    \includegraphics[width=\columnwidth]{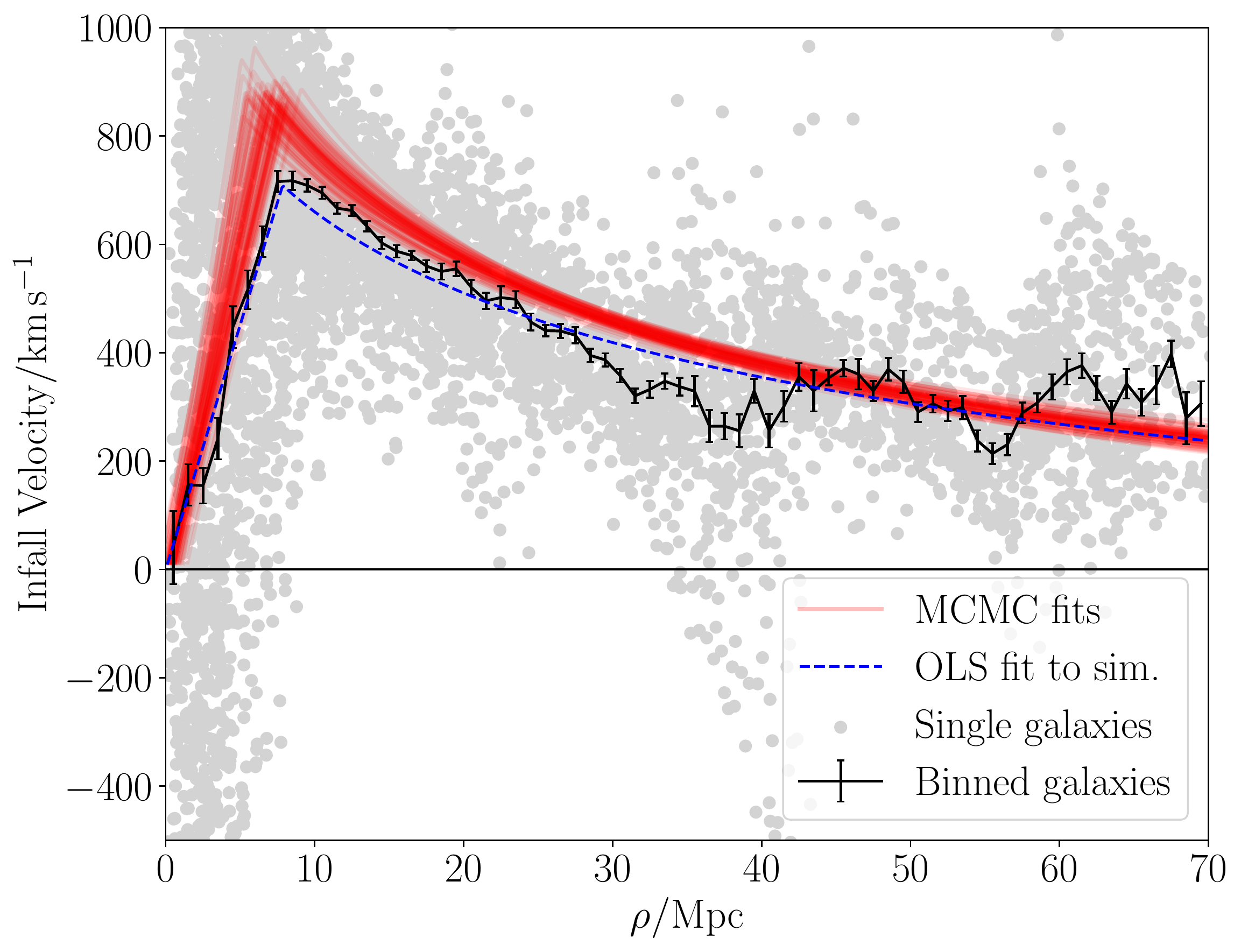}
    \caption{Example of MCMC fits to a single filament. One thousand semi-transparent fits drawn at random from the joint posterior distribution of all parameters (red lines) are overplotted on the simulation data (grey points) for individual galaxies from the observational wedge around the filament (before distance uncertainties were added). Considerable uncertainty exists near $\rho_\mathrm{max}$; however, by $\sim$25~Mpc all fits give consistent values for the infall velocity. The black errorbars indicate the mean and standard deviations in bins of width 1~Mpc for all galaxies in the simulation wedge (with no distance uncertainties added), and the blue dashed line is the ordinary least square fit to these binned data.}
    \label{fig:MCMC_overlay}
\end{figure}

\begin{figure}
    \centering
    \includegraphics[width=\columnwidth]{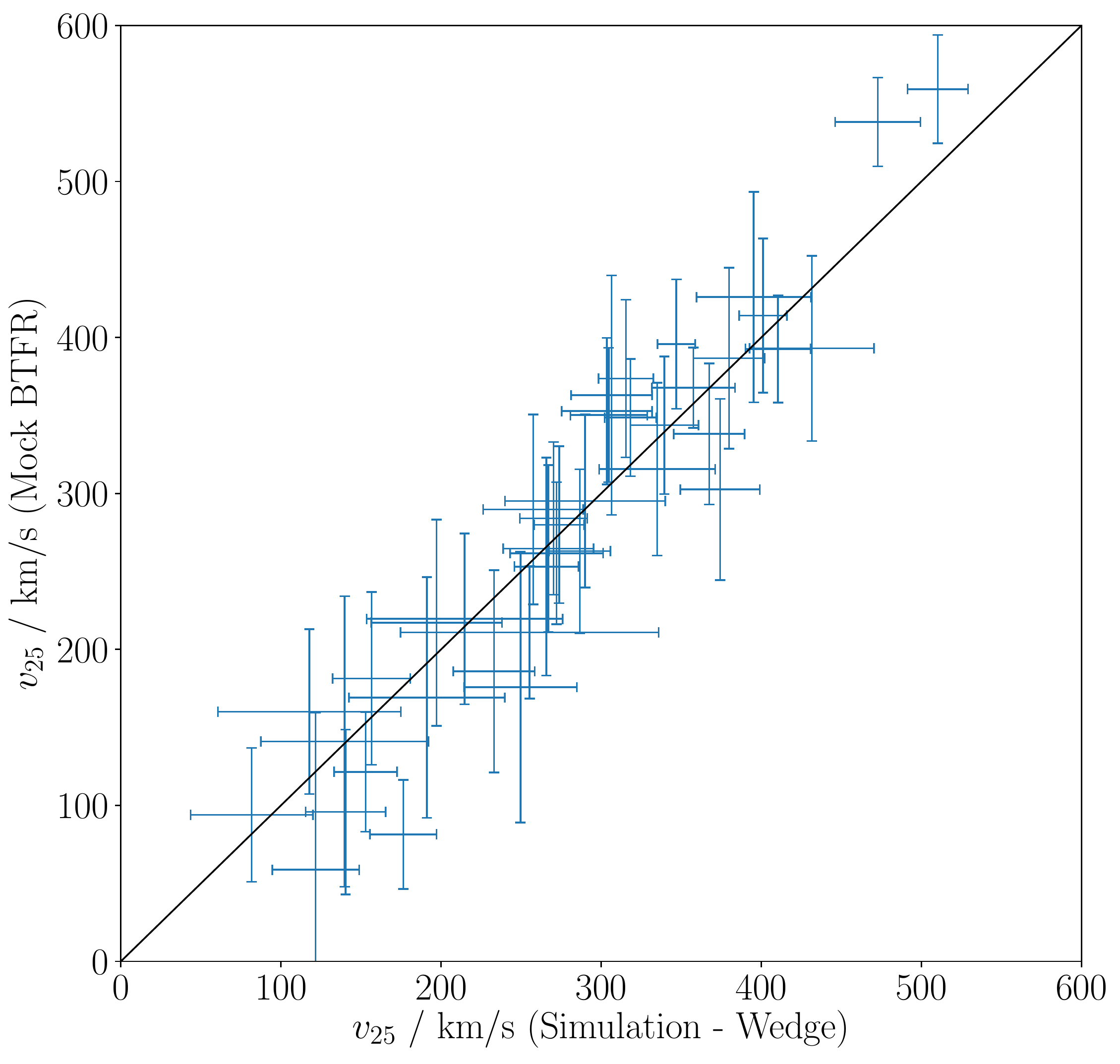}
    \caption{Infall velocity at 25~Mpc for observational wedges before  (x-axis) and after (y-axis) introducing the effects of BTFR-like distance uncertainties. Each point corresponds to a single PPS-like filament and each $V_{25}$ value (y-axis) was calculated from an MCMC fit of 2000 galaxies with 0.2~dex distance uncertainties (red lines in Figure \ref{fig:MCMC_overlay}). For each simulation wedge (without distance uncertainties) all galaxies were averaged in bins of width 200~kpc in radial distance from the principal axis. The infall function was then fitted via an OLS fit to the bin median values (dashed blue line in Figure \ref{fig:MCMC_overlay}). The uncertainties for the mock values represent propagation of the standard deviations of the posterior distributions (about the mean) of $\rho_\mathrm{max}$, $V_\mathrm{max}$, and $\rho_0$. For the simulation errorbars the same propagation of uncertainty was performed for the 1-$\sigma$ OLS fit uncertainties in these parameters.}
    \label{fig:v25_mock_slice}
\end{figure}

\begin{figure}
    \centering
    \includegraphics[width=\columnwidth]{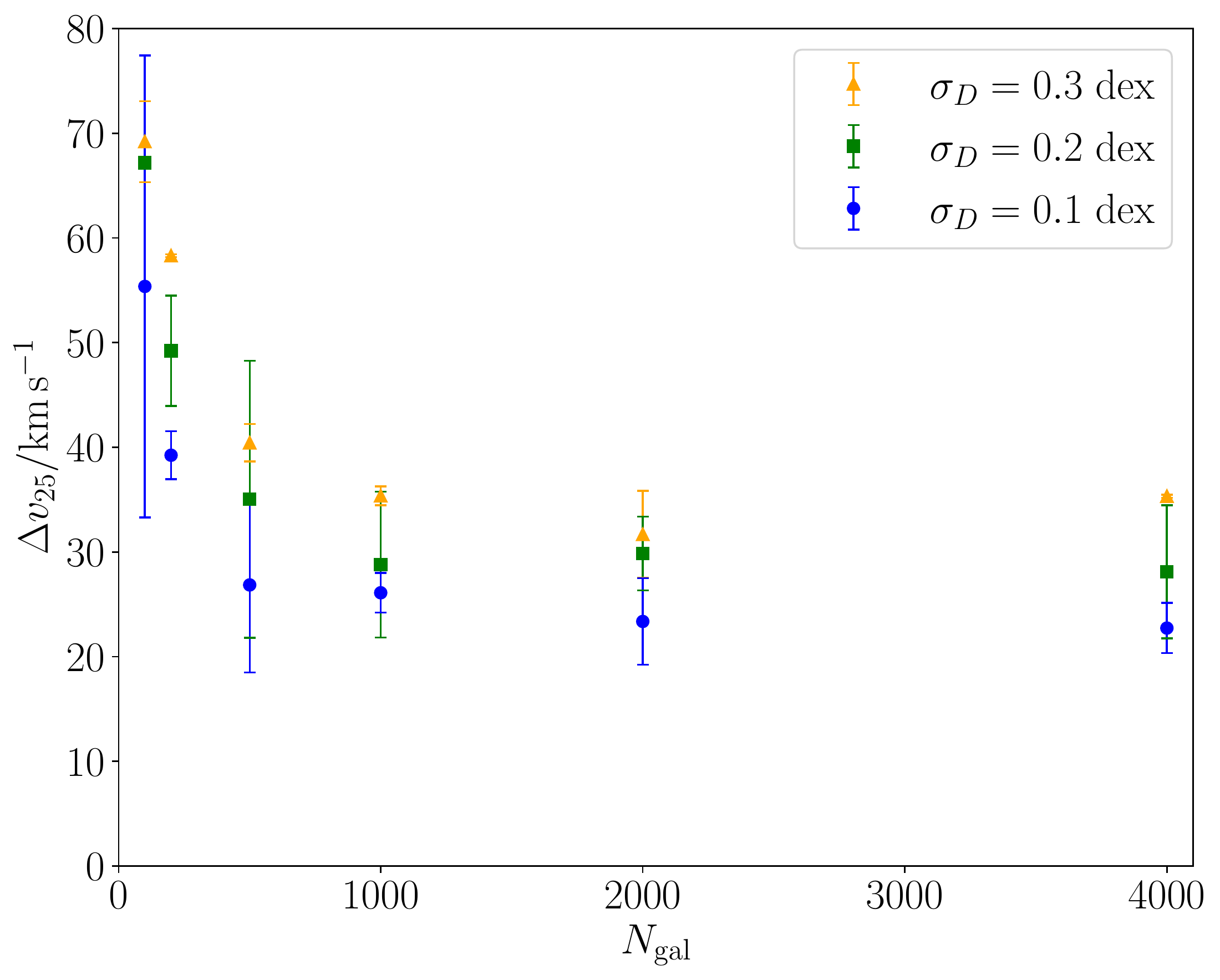}
    \caption{Median offset between the mock $V_{25}$ fit value and that of the simulation wedge with no distance uncertainties as a function of sample size and scale of distance uncertainties. Each point corresponds to the median value over 40 PPS-like filaments fit using MCMC.} Error bars correspond to jackknife standard deviation estimates for the 40 filaments fit in each data point.
    \label{fig:sample_size_err}
\end{figure}

\subsection{Constructing Mock Observations} 

In this analysis we construct wedge-shaped volumes around each simulated filament that are designed to approximately reflect the real geometry of APPSS. The filament is placed at a distance of 70~Mpc\footnote{We note that throughout this section we will use Mpc, to indicate physical distance with $h=0.7$.} and aligned perpendicular to the line-of-sight. The wedge is truncated at twice the distance to the filament, that is, at $z=140$~Mpc. The maximum dimensions of the wedge are $80\times40$~Mpc, such that a section of the filament 40~Mpc long is present in the middle of the wedge, and 10~Mpc above and below the filament (perpendicular to both the filament axis and the line-of-sight).  

Even if we assume the filament is oriented across the sky, it will not look the same from all viewing directions unless the filament and the region surrounding it are axially symmetric. In order to investigate the degree to which this introduces uncertainty, we create wedges that view each filament from eight equally spaced angles around the principal axis.

In order to mimic the APPSS data, we use a detection limit based on the cold atomic gas mass of each simulated galaxy. Based on the preliminary APPSS data \citep{odonoghue2019a} we expect the typical sensitivity to be approximately three times that of ALFALFA \citep{haynes2018a}. Therefore, to construct the mock completeness limit we took the ALFALFA 50\% completeness limit \citep{haynes2011} for a source with a velocity width of 50~\kms \ and decreased the flux by a factor of three.  This gave the following approximate completeness limit for an APPSS-like survey: $\log M_\mathrm{HI}/M_\odot \ge 2\log(D/\mathrm{Mpc})+ 4.54$. The real sensitivity of \HI \ observations also depends on the velocity width of each source, however, for this exercise the above approximate limit is sufficient as it requires mock galaxies to have significant cool gas reservoirs and enforces a flux-limited behaviour in the sample selection.  Figure 12 shows two sample mock observational wedges for a single filament, illustrating the galaxy sample and the degree to which its appearance depends on viewing angle.

In order to construct infall profiles from mock data, we convert observational coordinates into $\rho$ and v$_\rho$. Because the observations measure three position coordinates (Right Ascension, Declination, and distance), it is just a matter of geometry to find the coordinate $\rho$ (see Appendix A).  However, the observer measures only one velocity component:  the motion along the line of sight, as determined from the observed redshift and distance. Therefore, it is not possible to convert the velocity to cylindrical filament-based coordinates without making some assumption about the unknown velocity components. One approach is to simply use the line-of-sight peculiar velocity as a proxy for v$_\rho$, with the sign depending on whether the galaxy is on the near side or far side of the filament; this is reasonable to the extent that the galaxies are located in a thin, narrow volume that the filament cuts across. Another approach is to assume that the motion is directly along the coordinate $\rho$, and that the observed line-of-sight peculiar velocity is simply a component of this motion; an analogous approach is sometimes used to convert observational coordinates to spherical infall coordinates for clusters of galaxies \citep[e.g. ][]{karachentsev2010a}. 

Figure 13 shows the degree to which each of these approaches is able to reproduce the known, true value of v$_\rho$ for one specific mock observational wedge. In this figure, galaxies are color-coded by the geometrical correction factor that would be applied to convert the line-of-sight velocity to v$_\rho$ under the assumption that the motion is toward the filament (see Appendix A for the derivation of this factor). Both approaches, using the line-of-sight velocity v$_\mathrm{los}$ (left panel of Figure 13) or applying the correction factor (center panel of Figure 13) work fairly well for galaxies in most of the volume, where the correction factor is less than about 1.2 (colored blue in the figure). However, for galaxies above and below the filament, in the region with a high correction factor (colored yellow), values scatter widely from the known values of v$_\rho$, making it important to either drop these galaxies entirely or carefully model their uncertainties. (We take the latter approach below). This occurs because in this region the line-of-sight velocities of the galaxies are close to perpendicular to the radial infall direction. Thus, in practice v$_\mathrm{los}$ has a minimal component from the true v$_\rho$, and the large correction factors only act to amplify unrelated components of the galaxies' motions. About 20\% of the galaxies in the observational wedge are in the region where the correction factor is above 1.2. 

\begin{figure*}
   \includegraphics[width=1.0\textwidth]{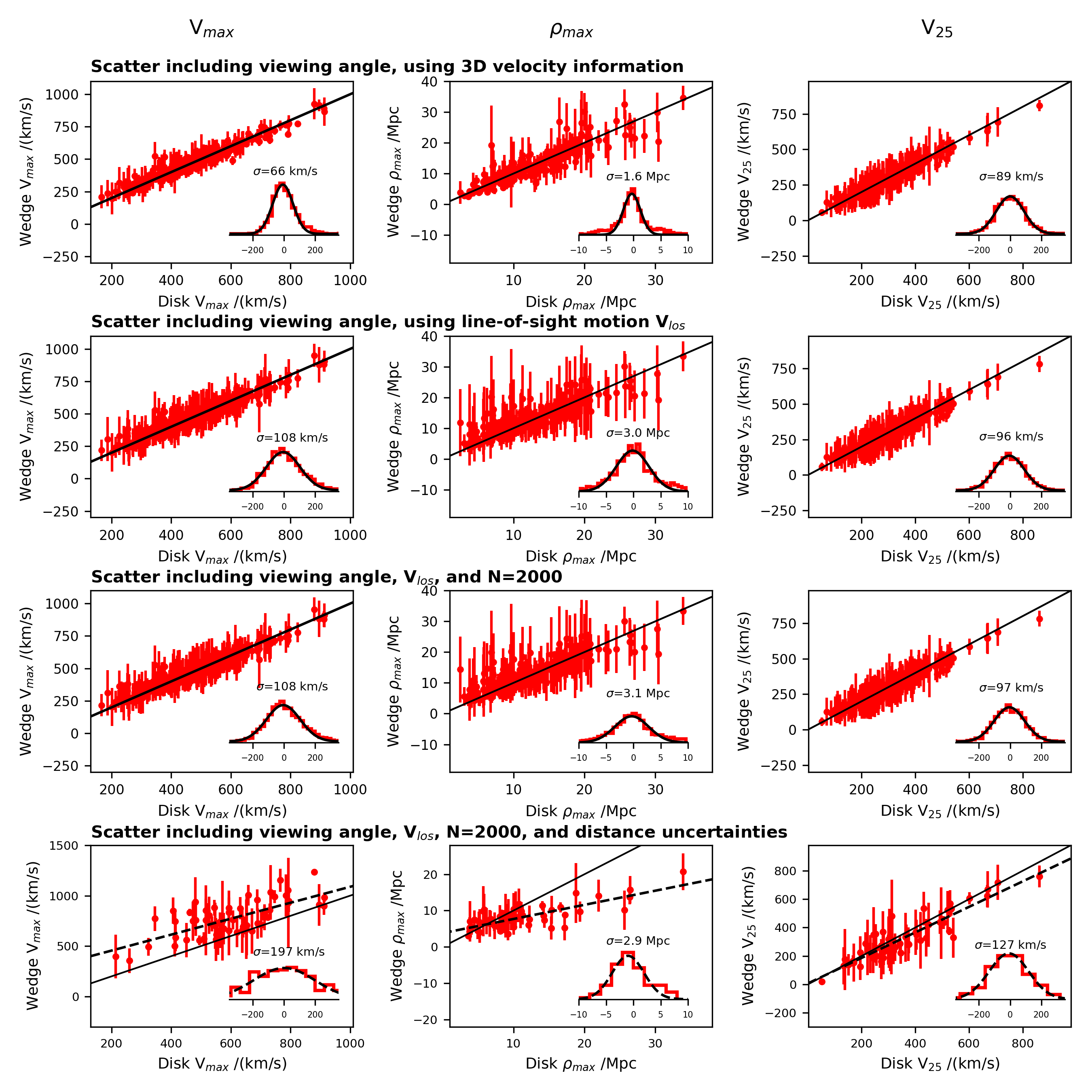}
    \caption{The scatter in $V_{max}$ (left column), $\rho_{max}$ (center column), and $V_{25}$ (right column) introduced by adding progressively more observational limitations: viewing angle (top row), velocity limited to line-of-sight component (second row), sample size limited to 2000 galaxies (third row), and distance uncertainties (bottom row).  Each panel shows the relationship between the mock observational wedges and the value obtained using a cylinder and full 3D information. Points show the average for the wedges for each filament as seen from different angles, with error bars showing the standard deviation. Solid lines show in each panel show the identity relation, and the dashed lines in the bottom row show the best fit once all uncertainties are included. Inset histograms show the deviation of the values for all the wedges from the 3D cylinder values (for the top three rows) and the deviation from the dashed best-fit line (for the bottom row). As more observational limitations are added, the deviation tends to increase.}
\end{figure*}

\begin{figure*}
   \includegraphics[width=1.0\textwidth]{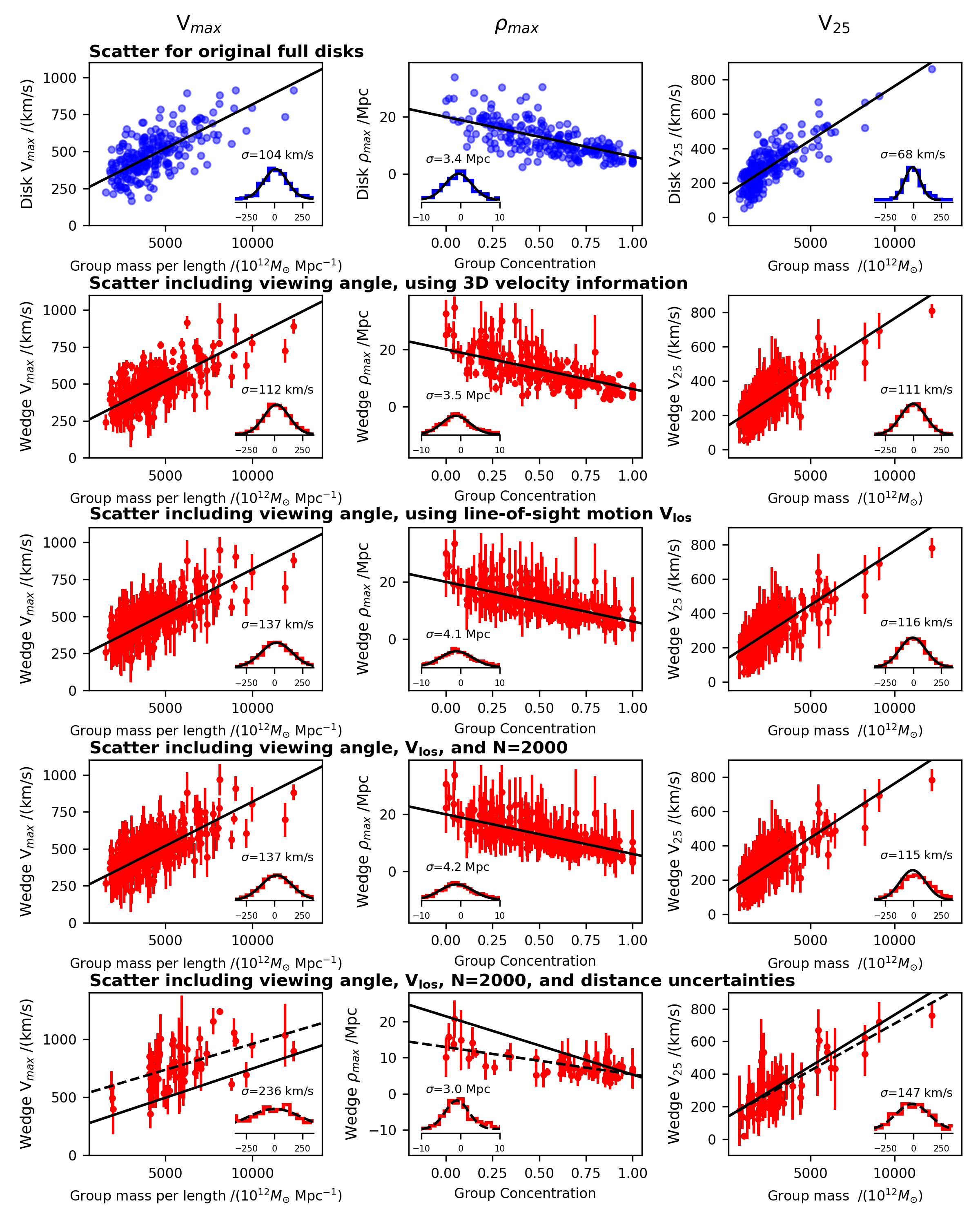}
    \caption{Scatter in relationships between $V_{max}$, $\rho_{max}$, and $V_{25}$ and the group properties of each filament, including the effects of observational limitations. The layout is similar to Figure 17, except that the top row shows the scatter for the original cylinders before introducing observational limitations. Solid lines show results from the fits to the original cylinders}.  Dashed lines in the bottom row show results from fits to the mock observations once all uncertainties are included. With all uncertainties included, it becomes especially difficult to recover the relationships for $V_{max}$, $\rho_{max}$.  The relationship for $V_{25}$ is more robust.
\end{figure*}

\begin{figure*}
   \includegraphics[width=1.0\textwidth]{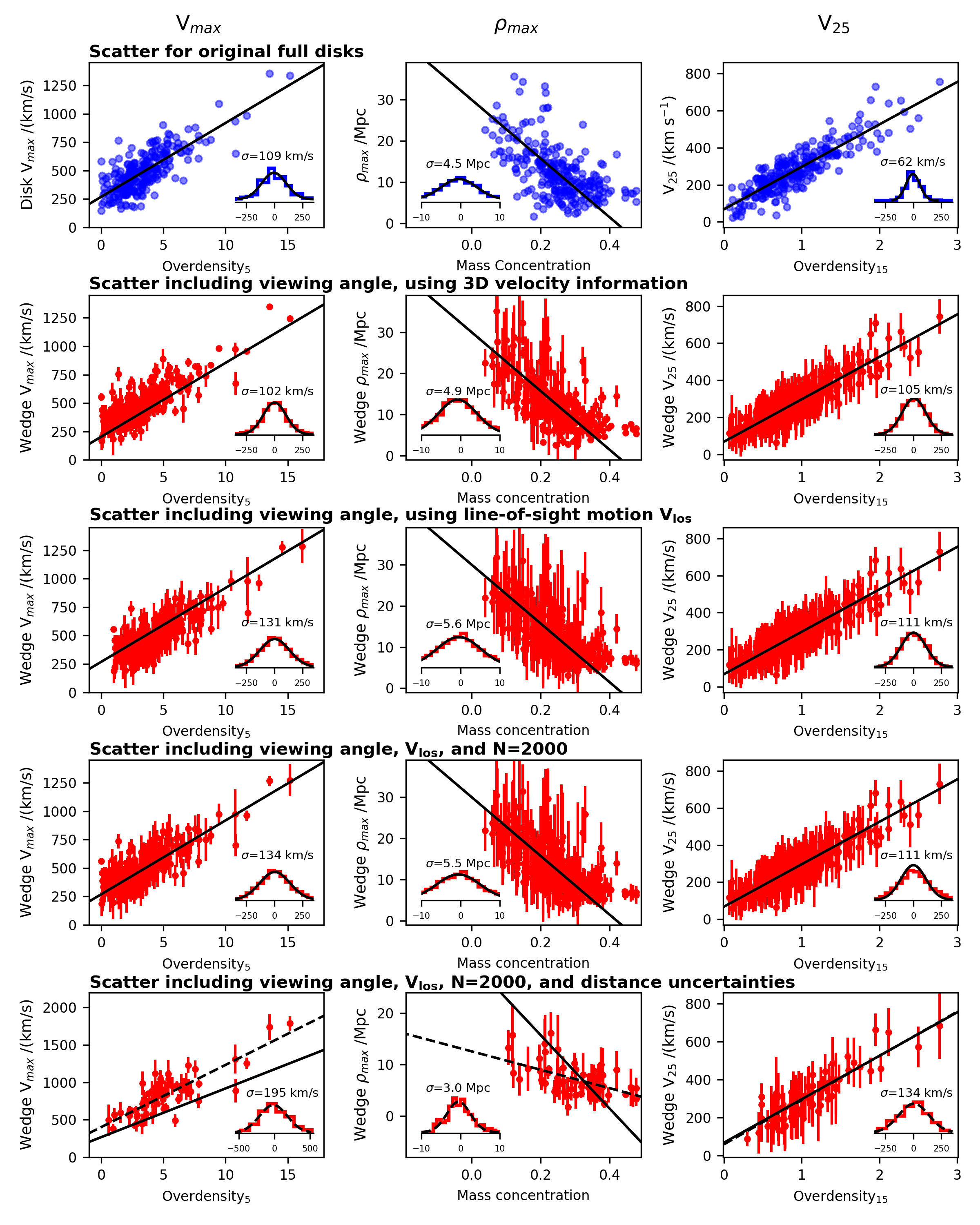}
    \caption{Scatter in relationships between $V_{max}$, $\rho_{max}$, and $V_{25}$ and the underlying matter distribution of filaments, including the effects of observational limitations. The layout is the same as Figure 18.}
\end{figure*}

Uncertainties are added to the distance values of the simulated galaxies similar to those expected for BTFR distance estimates. Specifically, random Gaussian errors are added to the distance in log-space (from a normal distribution with $\sigma=0.2$~dex), meaning the uncertainties are log-normally distributed and have a long tail towards larger distances. As the BTFR is a relation between baryonic mass and rotation velocity, and observed mass depends on distance squared, this corresponds to a vertical scatter in the BTFR of 0.4~dex (total scatter, not the intrinsic scatter). This may seem unduly pessimistic; however, many of the APPSS galaxies are expected to be low-mass galaxies, for which the BTFR has increased scatter. For simplicity, no uncertainties are added to either the angular position of each object or the radial velocity, both of which are typically negligible in observations in comparison to the distance uncertainty.

The next, and principal, step is to fit the infall function to these mock data. The large distance uncertainties ($^{+41}_{-26}$~Mpc at 70~Mpc) make this highly challenging. Simply binning the sample with distance would lead to a large fraction of galaxies being placed in incorrect bins and a smoothing out of the true infall pattern. Furthermore, although the positions of the galaxies can in theory be precisely converted to the filament-centric cylindrical coordinate $\rho$, in practice it is usually impossible to determine (independently of its observed velocity) with confidence whether any given galaxy in the vicinity of the filament is in front or behind it. These limitations rule out a conventional fitting approach, but can be appropriately addressed by the many realizations used in a Markov Chain Monte Carlo (MCMC) approach. We therefore adopt such a fitting approach using the \texttt{Stan} Python package \texttt{pystan}.\footnote{Stan Development Team. 2021. Stan Modeling Language Users Guide and Reference Manual, 2.28. \url{https://mc-stan.org}} This removes the need to bin the data as the enormous uncertainties in the distances can be incorporated into a hierarchical model that is used to fit the infall model to the full data set.

We fit each filament using a random sample of 2000 galaxies from the mock catalog produced following the steps above. This is intended to roughly match the number of tracers that the final APPSS sample is expected to have over an equivalent volume. For these computationally intensive MCMC fits, we focus on the PPS-like filaments in the MR7 simulation, performing MCMC fits to mock observations of the forty filaments with 5625 $\pm$ 1591 M$_\odot$ Mpc$^{-1}$ and group concentration between 0.6 and 1.0. We run MCMC fits for each of these filaments from four different viewing angles, separated by 90$^\circ$. In addition, we run MCMC fits on a smaller sample of twenty filaments (also from four viewing angles) that span the full range of filament properties. Finally, we also fit a sample of filaments from the MR run, using the same selection criteria, in order to allow comparison to the underlying density field that is unavailable for the MR7 run of the Millennium Simulation.

Our initial tests of this fitting approach demonstrated that given the large distance uncertainties it was unlikely to be able to reliably recover the form of the infall very close to the filament spine (e.g. for $\rho < 10$~Mpc). We therefore adopted a modified version of Equation \ref{eqn:2paraminfall} where an additional parameter, $\rho_0$, was added to allow the exponential decay in infall velocity to be fit independently of the slope of the inner linear rise:
\begin{equation}
    \label{eqn:3paraminfall}
    \begin{split}
    v = V_{\rm max}\  \frac{\rho}{\rho_{\rm max}}, \ \ \ \ \ \ \rho \leq \rho_{\rm max} \\
    v = V_{\rm exp}\ e^{-\sqrt{\frac{\rho}{\rho_{0}}}},  \ \rho > \rho_{\rm max}   
    \end{split}
\end{equation}
where $V_{\rm exp} = V_{\rm max} \exp{\sqrt{\rho_{\rm max}/\rho_0}}$.

The model parameters fit with the MCMC are those of this infall function plus the distance to the filament spine, the scale of random peculiar velocities (not associated with the filament), and the true distance to every galaxy in the sample. The last of these adds 2000 additional nuisance parameters which are marginalised at the end of the MCMC run. This greatly increases the complexity of the model, but it is necessary to address the enormous (and asymmetric) distance uncertainties that BTFR distance estimates entail. Due to this increase in parameters a normal desktop computer takes approximately 2~h to achieve convergence and fit a single filament.

We provide mostly broad priors for the model as follows: $V_\mathrm{max} \sim \mathrm{normal}(\mu=500\,\mathrm{km\,s^{-1}},\sigma=200\,\mathrm{km\,s^{-1}})$, $D_\mathrm{fil} \sim \mathrm{normal}(\mu=70\,\mathrm{Mpc},\sigma=0.5\,\mathrm{Mpc})$, $\rho_\mathrm{max} \sim \mathrm{normal}(\mu=5\,\mathrm{Mpc},\sigma=2\,\mathrm{Mpc})$, $\rho_0 \sim \mathrm{normal}(\mu=20\,\mathrm{Mpc},\sigma=10\,\mathrm{Mpc})$, $\sigma_\mathrm{V,pec} \sim \mathrm{normal}(\mu=200\,\mathrm{km\,s^{-1}},\sigma=50\,\mathrm{km\,s^{-1}})$. In addition to these priors, all parameters are constrained to be positive. The distance prior is fairly tightly constrained as there are so many galaxies in the real PPS filament that their average velocity can be used to estimate an accurate distance simply by assuming Hubble flow. The prior for the true distance to each galaxy is $p(D_\mathrm{true}|D_\mathrm{max}) = 3 D_\mathrm{true}^{2}/D_\mathrm{max}^{3}$, that is, assuming that the galaxies are distributed randomly throughout the volume out to $D_\mathrm{max} = 140$~Mpc. This prior does not include the overdensity of the filament itself (which may be a means to marginally improve the model), but it will account for Malmquist bias by including the prior knowledge that galaxies are likely, on average, to be further away than their distance estimates would imply.

An example of a fit to a particular mock filament is provided in Figure \ref{fig:MCMC_overlay}, which shows a random selection of fits from the MCMC chain on top of points showing the infall velocity of each galaxy in the sample as a function of $\rho$. Note that these points correspond to galaxies before distance uncertainties are added, as if the uncertainties were included then no trend would be visible by eye. Appendix B provides a corner plot for this filament, showing posterior distributions of the model parameters.

The final step in the MCMC fitting process is to calculate $V_{25}$. This is found by evaluating the expression for the infall function at $\rho=25$~Mpc using the mean fit parameters from the MCMC posterior. Errors are estimated by propagating the uncertainties (one-$\sigma$ standard deviation of posteriors). Figure \ref{fig:v25_mock_slice} compares these $V_{25}$ values to those calculated from an ordinary least-squares (OLS) fit to all galaxies in the same mock wedges (before distance uncertainties were added and the samples were reduced to 2000 galaxies), binned into 0.1~Mpc wide bins in $\rho$. We see that overall there is excellent agreement (within the uncertainties) in the $V_{25}$ values from the MCMC fits and the OLS fits (with no distance uncertainties). The tight correlation between the two measurements of $V_{25}$ demonstrates that the MCMC fitting procedure is able to accurately recover this quantity even with the considerable uncertainties of BTFR distance estimates.

We expand further on this test in Figure \ref{fig:sample_size_err} where we plot the median deviation between these two fits (essentially the scatter about the one-to-one line in Figure \ref{fig:v25_mock_slice}) as a function of both the number of tracers used in the MCMC fits and the scale of the BTFR distance uncertainties that we assume. The behaviour is broadly as expected, with the typical size of deviations decreasing as more tracers are used in the fits and as the assumed BTFR distance uncertainties decrease. For small sample sizes ($N<500$) the accuracy of the fits rapidly improves with increasing sample size, but beyond $N=2000$ there is very little benefit to expanding the sample of tracers. The improvements from reducing the scale of the distance uncertainties can be seen for all sample sizes; however, it is most pronounced for small samples.

\subsection{Results from Mock Observations} 

Each source of observational uncertainty has the potential to introduce challenges to measuring $V_{\rm max}$, $\rho_{\rm max}$, and $V_{25}$. Figure 17 summarizes how well we can recover the values of these three parameters as we include progressively more sources of uncertainty. The top row shows the scatter that is introduced by observing the filament from a specific viewing angle, taking into account galaxy sample selection based on HI flux as described above. Most of the observational wedges include about 5000 galaxies above this limit. The values obtained from the wedge-shaped observational volumes are compared with the values obtained using the full, three-dimensional information from galaxies in the cylinders, as described in Section 3. For example, the top left panel shows the average value for $V_{\rm max}$ obtained from the eight observational wedges that view the filament from different angles around the principal axis, with error bars showing the standard deviation. These are plotted against the values for $V_{\rm max}$ obtained using the three-dimensional cylinders. The straight line indicates the identity relation, and the inset histogram shows the scatter around the identity relation. The histogram includes each of the viewing angles for each filament as separate data points, and shows the best-fit Gaussian curve and the standard deviation. For $V_{\rm max}$, for example, the values for observational wedges have a Gaussian distribution about the identity relation with a scatter of 66 km s$^{-1}$. 

The second row of Figure 17 includes an additional source of uncertainty: the fact that only the line-of-sight motion $V_{\rm los}$ is available in the observations. To do this, we consider the results from our attempts to geometrically recover $V_{\rho}$ from $V_{\rm los}$ combined with position information.  As shown in Figure 13, $V_{\rm los}$ and $V_{\rho}$ are very similar for galaxies in most parts of the observational wedge but not for galaxies just above and below the filament, where only a small component of $V_{\rho}$ is along the line of sight. About 20\% of the galaxies in the wedge are typically in the region where the geometrical correction factor is greater than 1.2. The uncertainty in $V_{\rho}$ is so large for these galaxies, whether or not we attempt a correction using the geometrical correction factor in Figure 13, that we remove them from the sample. Note that the option of removing galaxies that are in this region is only feasible if their distances can be determined with precision. In the context of the second row of Figure 18, we use this approach as a way to estimate the scatter introduced specifically by the lack of three dimensional velocity information when combined with the scatter from using observational wedges. (However, when we include distance uncertainties using MCMC fits, we include all the galaxies.) For example, the scatter in $V_{\rm max}$ rises to 108 kms$^{-1}$ in the second row, compared with 66 km s$^{-1}$ in the first row.  If the uncertainties add in quadrature, this implies an additional scatter of $\sqrt{108^2 - 66^2}$ kms$^{-1}$ = 85~kms$^{-1}$ specifically from the observational limitation to one velocity component.  

The third row of Figure 17 combines the limitations from the first two rows with a limited sample size of 2000 galaxies, down from about 5000 that are above the HI flux limit for most filaments. This adds only a small increase in scatter -- for example, small enough that the scatter in $V_{\rm max}$ is still about 108 kms$^{-1}$. 

The fourth row of Figure 17 includes all the uncertainties in our mock observations, including distance uncertainties of 0.2~dex, taken into account using MCMC fits. Here we see significant limitations to the two-parameter description of infall in terms of $V_{\rm max}$ and $\rho_{\rm max}$. The scatter for $V_{\rm max}$ is both large and shifted to higher values, while the values for $\rho_{\rm max}$ mock observations match only for values less than about 12 Mpc; otherwise they are considerably underestimated. Using a one-parameter description of infall in terms of $V_{\rm 25}$ is more promising. In terms of the adjusted coefficient of determination $R^2_{adj}$, the best-fit (dashed) lines for $V_{\rm max}$, $\rho_{\rm max}$, and $V_{\rm 25}$ have values of 0.237, 0.278, and 0.485, respectively. 

We next consider the possibility of predicting the infall rate based on the structure of the filament. As we showed in the top rows of Figure 10 and 11, $V_{\rm max}$, $\rho_{\rm max}$, and $V_{25}$ are related to the group masses and group concentration of the filaments. Figure 18 shows how the scatter in these relationships increases as we introduce observational uncertainties. The top row shows the original scatter in these relationships for the full cylinders, and the following rows introduce each type of observational uncertainty, using the same format as in Figure 17. The bottom row, which includes all the uncertainties, reflects the same limitations seen in the bottom row of Figure 16.  For $V_{\rm max}$ the scatter is high and the values are offset from the original relation, and for $\rho_{\rm max}$ the values are offset and the slope is flatter. As expected, a more promising approach in the context of our full sample of simulated filaments is to use $V_{\rm 25}$. In terms of the adjusted coefficient of determination $R^2_{adj}$, the best-fit (dashed) lines for $V_{\rm max}$, $\rho_{\rm max}$, and $V_{\rm 25}$ have values of 0.135, 0.213, and 0.396, respectively. Taking into account all the sources of uncertainty, filaments that are ``like the PPS" in their observable group structure, as defined in Section 3, have an average $V_{\rm max}=791$ km s$^{-1}$ with standard deviation 230 km s$^{-1}$, an average $\rho_{\rm max}=7.0$ Mpc with standard deviation 3.5 Mpc, and an average $V_{\rm 25}=301$ km s$^{-1}$ with standard deviation 150 km s$^{-1}$.

Finally, we consider the relationships between $V_{\rm max}$, $\rho_{\rm max}$, and $V_{25}$ and the underlying mass distribution.  As we showed in the bottom rows of Figure 10 and 11, $V_{\rm max}$, $\rho_{\rm max}$, and $V_{25}$ are related to the  mass and mass concentration of the filaments, including dark matter. Figure 19 shows how the scatter in these relationships increases as we introduce observational uncertainties. In terms of the adjusted coefficient of determination $R^2_{adj}$, the best-fit (dashed) lines for the fits that include all uncertainties (the bottom row) $V_{\rm max}$, $\rho_{\rm max}$, and $V_{\rm 25}$ have values of 0.570, 0.153, and 0.427, respectively. In this case, the relationship for $V_{\rm max}$ shows the most predictive value and the relationship for $\rho_{\rm max}$ shows the least predictive value. 

\section{Summary and Discussion}

In this paper, we characterize the theoretically expected large-scale infall pattern of galaxies toward cosmological filaments, and we model the uncertainties in determining the infall profile using measurements like those in the Arecibo Pisces-Perseus Supercluster Survey (APPSS).  Our results can be summarized as follows:

\begin{itemize}

\item The average infall profile, when stacked by distance from the principal axis, follows the equation $v_{\rho} = C\rho^4 e^{{-\rho}^{ \ 0.2}}$. The rounded peak in this function is the result of averaging together filaments with velocity profiles that peak at different distances. When the distance for each filament is normalized by its peak infall distance the average profile has a sharper peak, and is better fit by the piecewise function given by Equation 2. This piecewise function is the appropriate one to use in fitting the profile of an individual filament.

\item The infall for each filament can be characterized by the maximum infall velocity $V_{\rm max}$ and the distance of the point of maximum infall from the principal axis of the filament $\rho_{\rm max}$. These two parameters are related to the properties of the groups in the filament: $V_{\rm max}$ is positively correlated with the mass in halos per length along the filament, and $\rho_{\rm max}$ is negatively correlated with the degree to which the halos are concentrated toward the principal axis. Based on these constraints, we expect the PPS infall region have $V_{\rm max}= 612\pm 116$ km s$^{-1}$ and $\rho_{\rm max}=8.9\pm 2.1$ Mpc (assuming h=0.7).

\item A simpler way to characterize infall, which is easier to determine observationally, is in terms of the single parameter $V_{25}$, the infall at a distance of 25 Mpc from the axis of the filament. Based on the mass of the central groups in the PPS, we predict $V_{25} = 329 \pm 68$ km s$^{-1}$.

\item The values for $V_{\rm max}$, $\rho_{\rm max}$, and $V_{25}$ also reflect the underlying mass distribution in the filament, including dark matter, so they have the potential to be used to constrain the large-scale dark matter distribution. The specific relationships between these parameters and the mass distribution are given in Figures 10 and 11. 

\item Our results are robust to the small differences in cosmological parameters between the MR7 and MR runs, and to the choice of selecting galaxies by cold gas mass or stellar mass. 

\item A significant source of uncertainty for observational measurements is the viewing angle of the filament, even if we assume the filament is oriented across the sky. For example, this introduces a scatter of 66 km s$^{-1}$ in the relationship between $V_{\rm max}$ as determined using a 3D cylinder versus that obtained from observational wedges for the same filament.

\item An additional source of uncertainty in determining the infall profile from observational data is the availability of only one velocity component, the motion along the line of sight. This is primarily a problem for galaxies in the region near the filament where the line of sight is nearly perpendicular to the direction of infall. For example, if we include only galaxies with velocity correction factors less than 1.2, then including this source of uncertainty along with the uncertainty in viewing angle increases the scatter in $V_{\rm max}$ to about 108 km s$^{-1}$.

\item The galaxy sample size introduces a significant source of additional uncertainty only if the sample size is less than about 2000. 

\item A particular challenge in measuring the infall profile is introduced by the uncertainty in galaxy distances, which can be approximately 30\% using the BTFR. We model this uncertainty using Markov chain Monte Carlo sampling, and find that it is particularly difficult to recover the expected values for $V_{\rm max}$ and $\rho_{\rm max}$, while recovering the single parameter $V_{\rm 25}$ is more promising.

\item Taking into account all sources of observational uncertainty, we find that of the three parameters $V_{\rm max}$, $\rho_{\rm max}$, and $V_{25}$ , it is $V_{25}$ that has the tightest correlation to the observable group structure of the filament, and $V_{\rm max}$ that has the tightest correlation with the underlying mass distribution. 

\item Taking into account the measurement uncertainties we expect for APPSS, we expect our measurements of the three parameters for the PPS to be in the ranges $V_{\rm max} = 791 \ \pm$ 230 km s$^{-1}$, $\rho_{\rm max} = 7.0 \pm  3.5$ Mpc, and $V_{25} =301 \pm$ 150 km s$^{-1}$. 

\end{itemize}

We emphasize that in this context we use a definition of filaments that is straightforward to apply to observations like those available for the PPS, and therefore does not take into account the detailed structure that can be seen in simulations.  For example, we determine the axis of the filament using principal component analysis on cluster-sized halos, resulting in a simple, straight-line estimate of the axis. This differs from studies that trace the curves of a filament using more fine-grained measures of density. If we did this, we would be more sensitive to infall on smaller spatial scales, including the infall to individual halos, rather than the average infall on larger scales.

Although this study was motivated by the APPSS project, which focuses on one particular nearby filament, it would be especially interesting to measure the average infall profile for a sample of multiple filaments. One advantage of such a study, is that it would allow us to test the average infall profile in Equation 2, since we expect the data from one filament alone to have too much scatter from the average profile to provide an effective test.  Another advantage would be that it would allow us to test the the relationships we obtain between $V_{\rm max}$, $\rho_{\rm max}$, and $V_{25}$ and the structure of the filaments.  

\section{Acknowledgements} \label{sec:Acknowledgements}

We are grateful for the contributions of all members of the Undergraduate ALFALFA Team, especially Trevor Viscardi, Martha Haynes, and Rebecca Koopmann. We also thank John Helly for assistance in accessing the Millennium database. This work has been supported by NSF grants AST-1211005, AST-1637339 and AST-1637271.  


\bibliography{mybib}

\appendix
\section{Derivation of infall profiles from mock observations}

Here we show the derivation for converting positions and velocities from mock observations into infall profiles. The mock observations can be described using spherical coordinates with the observer at the origin. The observer measures three position coordinates: the distance $r$, and the angular positions on the sky $\theta$ and $\phi$. 
We can express the observer's coordinates in terms of Cartesian coordinates using
\begin{eqnarray}
    x &=& r \sin{\theta}\cos{\phi} \\
    y &=& r \sin{\theta}\sin{\phi} \\
    z &=& r \cos{\theta}. 
\end{eqnarray}
If the filament is positioned parallel to the $y$ axis at $x=D_\mathrm{fil}$ and $z=0$, the distance of a given point from the principal axis of the filament can be written as
\begin{equation}
    \rho = \sqrt{(x - D_\mathrm{fil})^2 + z^2}.
\end{equation}
This is similar to the usual cylindrical coordinate $\rho$, except that it is perpendicular to the $y$ axis instead of the $z$ axis, and the $x$ axis is shifted by  $D_\mathrm{fil}$.
Substituting expressions A1 and A3 into expression A4 gives $\rho$ in terms of the observational coordinates: 
\begin{equation}
    \rho = \sqrt{r^2 \cos^2{\theta} + (r \sin{\theta}\cos{\phi} - D_\mathrm{fil})^2}.
\end{equation}

Because the observer can measure three independent position coordinates, it is just a matter of geometry to obtain expression A5. On the other hand, the observer can measure only one velocity component: the motion along the line of sight $v_{r}$. Therefore, it is not possible to convert the velocity to cylindrical filament-based coordinates without making some assumption about the unknown velocity components. Here we take the approach of assuming that the motion is directly along the coordinate $\rho$; an analogous approach is sometimes used to convert observational coordinates to spherical infall coordinates for clusters of galaxies \citep[e.g. ][]{karachentsev2010a}. In this case, we can relate the line-of-sight component of the velocity $v_{r}$ to the motion toward the filament $v_{\rho}$ via the dot product
\begin{equation}
    v_{r} = \mathbf{v} \cdot \hat{\mathbf{r}} = v_\rho \hat{\boldsymbol{\rho}} \cdot \hat{\mathbf{r}}
\end{equation}
where 
\begin{equation}
     \hat{\mathbf{r}} = \frac{x}{r}\hat{\mathbf{x}} + \frac{y}{r}\hat{\mathbf{y}} + \frac{z}{r}\hat{\mathbf{z}}
\end{equation}
and 
\begin{equation}
     \hat{\boldsymbol{\rho}} = \frac{x-D_\mathrm{fil}}{\rho}\hat{\mathbf{x}} + \frac{z}{\rho} \hat{\mathbf{z}}.
\end{equation}
Writing A7 and A8 in terms of the observer's spherical coordinates using A1-A3, and taking the dot product,
\begin{equation}
     \hat{\boldsymbol{\rho}} \cdot \hat{\mathbf{r}} = \frac{\sin{\theta}\cos{\phi}(r\sin{\theta}\cos{\phi} -D_\mathrm{fil}) + r\cos^2{\theta} } {\sqrt{(r \sin{\theta}\cos{\phi} - D_\mathrm{fil})^2 +
     r^2 \cos^2{\theta}}}.
\end{equation}
Finally, solving for $v_{\rho}$ in A6 and substituting A9 for the dot product, we obtain
\begin{equation}
     v_\rho = v_{r} \frac {\sqrt{(r \sin{\theta}\cos{\phi} - D_\mathrm{fil})^2 + r^2 \cos^2{\theta}}} {\sin{\theta}\cos{\phi}(r\sin{\theta}\cos{\phi} -D_\mathrm{fil}) + r\cos^2{\theta} }.
\end{equation}
This provides the expression for the motion toward the filament in terms of the observational coordinates. The factor by which $v_r$ is multiplied to obtain $v_{\rho}$ is the geometrical correction factor illustrated in Figure 13.

\section{Sample MCMC fit}

Here we provide an example of our Markov Chain Monte Carlo (MCMC) sampling method applied to a particular mock filament. Figure \ref{fig:corner_plot} shows a corner plot of the joint posterior distribution of the 5 model parameters. This mock filament is the same one for which a random selection of fits from the MCMC chain are provided in Figure \ref{fig:MCMC_overlay}. 

\begin{figure}
    \centering
    \includegraphics[width=\columnwidth]{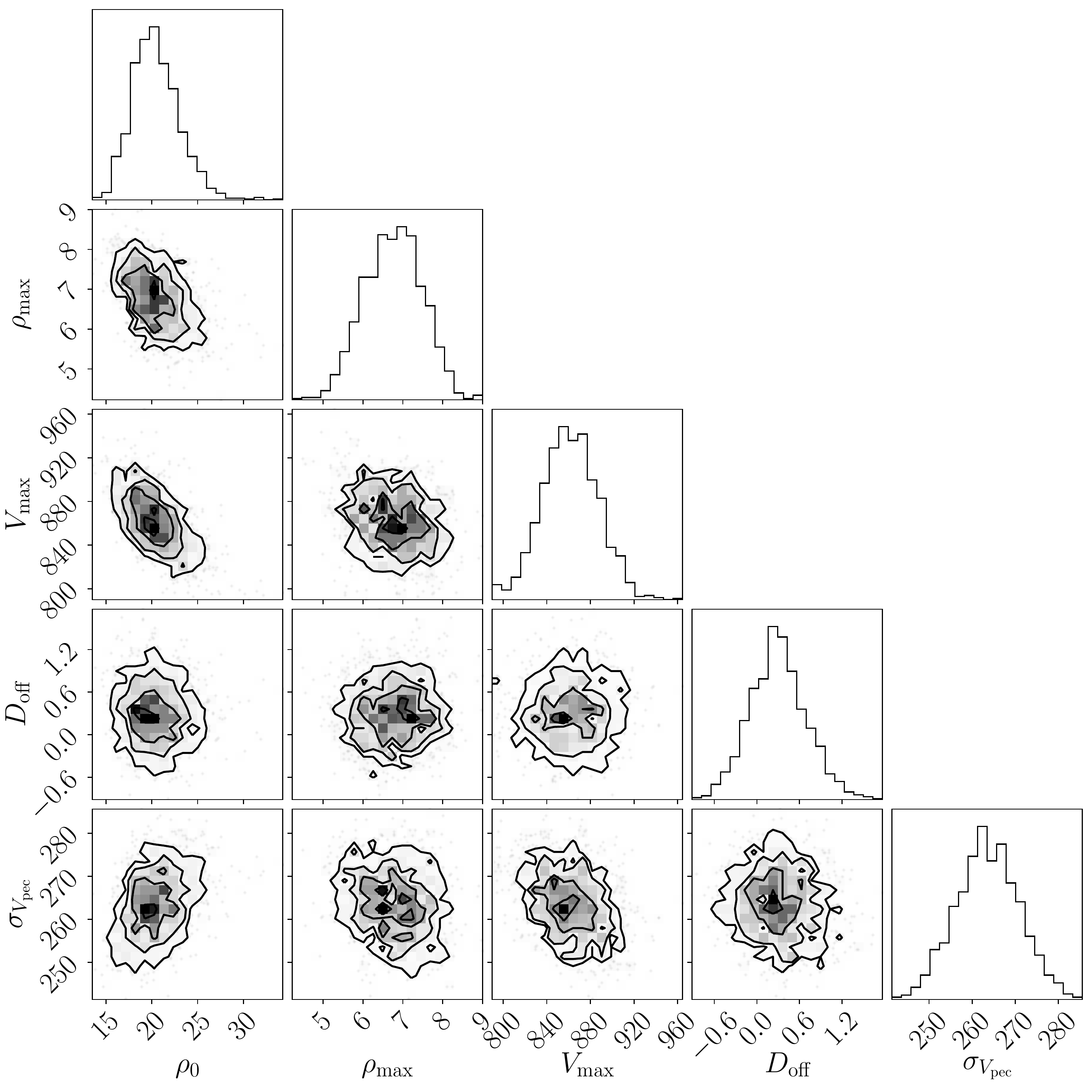}
    \caption{Example corner plot for one filament showing the posterior distributions of the model parameters fit with MCMC. In addition, to the three model parameters ($\rho_0$, $\rho_\mathrm{max}$, and $V_\mathrm{max}$) the MCMC fit also fits for any offset in the distance to the filament spine ($D_\mathrm{off}$) and fits for the typical scale of peculiar velocities (not associates with infall on to the filament), $\sigma_{V_\mathrm{pec}}$.}
    \label{fig:corner_plot}
\end{figure}

\end{document}